# Predictions for Observable Atmospheres of Trappist-1 Planets from a Fully Coupled Atmosphere–Interior Evolution Model

J. Krissansen-Totton and J. J. Fortney
Department of Astronomy and Astrophysics, University of California, Santa Cruz, Santa Cruz, CA 95064, USA; jkt@ucsc.edu
Received 2022 January 13; revised 2022 April 14; accepted 2022 April 19; published 2022 July 7

## Abstract

The Trappist-1 planets provide a unique opportunity to test the current understanding of rocky planet evolution. The James Webb Space Telescope is expected to characterize the atmospheres of these planets, potentially detecting $CO_2$, CO, $H_2O$, $CH_4$, or abiotic $O_2$ from water photodissociation and subsequent hydrogen escape. Here, we apply a coupled atmosphere–interior evolution model to the Trappist-1 planets to anticipate their modern atmospheres. This model, which has previously been validated for Earth and Venus, connects magma ocean crystallization to temperate geochemical cycling. Mantle convection, magmatic outgassing, atmospheric escape, crustal oxidation, a radiative-convective climate model, and deep volatile cycling are explicitly coupled to anticipate bulk atmospheres and planetary redox evolution over 8 Gyr. By adopting a Monte Carlo approach that samples a broad range of initial conditions and unknown parameters, we make some tentative predictions about current Trappist-1 atmospheres. We find that anoxic atmospheres are probable, but not guaranteed, for the outer planets; oxygen produced via hydrogen loss during the pre-main sequence is typically consumed by crustal sinks. In contrast, oxygen accumulation on the inner planets occurs in around half of all models runs. Complete atmospheric erosion is possible but not assured for the inner planets (occurs in 20%–50% of model runs), whereas the outer planets retain significant surface volatiles in virtually all model simulations. For all planets that retain substantial atmospheres, $CO_2$-dominated or $CO_2$–$O_2$ atmospheres are expected; water vapor is unlikely to be a detectable atmospheric constituent in most cases. There are necessarily many caveats to these predictions, but the ways in which they misalign with upcoming observations will highlight gaps in terrestrial planet knowledge.

*Unified Astronomy Thesaurus concepts:* Exoplanet atmospheres (487); Planetary atmospheres (1244); Astrobiology (74); Planetary interior (1248); Exoplanet atmospheric composition (2021); Exoplanet astronomy (486)

## 1. Introduction

With the recent launch of the James Webb Space Telescope (JWST), atmospheric constraints on terrestrial exoplanet atmospheres are imminent. The Trappist-1 system (Gillon et al. 2017; Luger et al. 2017), provides seven favorable targets for comparative planetology due to the star's small size and proximity to Earth (Turbet et al. 2020a). Transit observations with the Hubble Space Telescope (HST) and the Spitzer Space Telescope have refined the approximately Earth-like densities of all Trappist-1 planets (Agol et al. 2021), and transit spectroscopy strongly disfavors $H_2$-rich, primary atmospheres (de Wit et al. 2018; Moran et al. 2018). However, JWST will provide the first opportunity to detect and characterize high mean molecular weight secondary atmospheres in this system, and therefore test prevailing theories of rocky planet formation and evolution.

While the on-orbit performance of JWST (Schlawin et al. 2021) and stellar contamination from Trappist-1 photosphere nonhomogeneities (Ducrot et al. 2018; Rackham et al. 2018) are challenging to anticipate, simulated observations and retrievals imply that Trappist-1 atmospheres ought to be accessible by coadding transits (Morley et al. 2017). Indeed, simulated transit spectroscopy of Trappist-1 planets suggests a wide variety of secondary atmospheres may be detectable with a modest number of transits (Lustig-Yaeger et al. 2019a). While biological quantities of atmospheric oxygen (∼0.2 bar) will likely not be detectable for the Trappist-1 planets within the nominal mission lifetime of JWST (Fauchez et al. 2020b; Krissansen-Totton et al. 2018b; Wunderlich et al. 2019), redox evolution may be constrained via the detection (or nondetection) of near-IR (NIR) $O_2$–$O_2$ collisionally induced absorption features for dense $O_2$ atmospheres produced by extreme hydrogen loss (Lincowski et al. 2018; Lustig-Yaeger et al. 2019a). Atmospheric $CO_2$ is also likely detectable in a small number of transits (Krissansen-Totton et al. 2018b; Wunderlich et al. 2019, 2020; Gialluca et al. 2021), and while water vapor on temperate planets probably condenses below transit chords (Fauchez et al. 2019; Komacek et al. 2020), steam-rich atmospheres from runaway greenhouse climates could be detectable (Lustig-Yaeger et al. 2019a). It may even be possible to observe the isotopic fractionation due to extreme hydrogen loss (Lincowski et al. 2019).

If these expected observational capabilities are realized, JWST could determine whether the Trappist-1 planets have retained steam atmospheres, are desiccated Venus-like objects with $CO_2$-rich atmospheres, or whether they are instead airless rocks due excessive atmospheric erosion via thermal (Bolmont et al. 2017; Bourrier et al. 2017) and nonthermal (Dong et al. 2018; Garcia-Sage et al. 2017; Kral et al. 2018) processes under the harsh radiation (and high impact velocity) environment of Trappist-1. Distinguishing between hazy, high mean molecular weight atmospheres and airless rocks may require many transits (Lustig-Yaeger et al. 2019b). For the habitable zone planets in the system, (e–g), temperate $N_2$–$CO_2$–$CH_4$ atmospheres may also be characterizable (Krissansen-Totton et al. 2018b; Lustig-Yaeger et al. 2019a; Wunderlich et al. 2019, 2020).

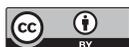

In addition to constraining bulk atmosphere composition, understanding planetary redox evolution via Trappist-1 observations is of strong astrobiological interest. Oxygen has long been considered a promising exoplanet biosignature (Meadows et al. 2018), but numerous hypothetical scenarios have been proposed for the build-up of abiotic oxygen on terrestrial planets due to secular redox evolution divergent from that of the Earth. The extended pre-main sequence of low-mass stars, such as Trappist-1, exposes main-sequence habitable zone planets to hundreds of millions of years of high bolometric and X-ray and ultraviolet (XUV) fluxes, potentially resulting in extensive hydrogen loss and abiotic oxygen accumulation (Bolmont et al. 2017; do Amaral et al. 2022; Luger & Barnes 2015). The balance between oxygen buildup via such hydrogen escape and oxygen loss via magma ocean oxidation has been explored in coupled models of thermal escape and magma ocean evolution (Schaefer et al. 2016; Wordsworth et al. 2018), including for Trappist-1e, f, and g specifically (Barth et al. 2021). While these models provide insights into the likelihood of pre-main sequence, abiotic oxygen accumulation, they only consider hydrogen and oxygen cycling during the magma ocean itself, and mostly do not explore how terrestrial atmospheres evolve after magma ocean crystallization, including the impact of crustal oxygen sinks. Moreover, such models do not consider how carbon cycle feedbacks may modulate planetary redox evolution.

The Planetary Atmosphere, Crust, and MANtle (PACMAN) evolution model simulates terrestrial planet evolution from magma ocean origins through to temperate, solid-state geochemical cycling (Krissansen-Totton et al. 2021a, 2021b). The model explicitly calculates mantle and atmospheric redox evolution due to hydrogen escape, magma ocean crystallization, and crustal oxygen sinks such as magmatic degassing, water–rock reactions, and direct oxidation of new crust. The model has been validated for both Earth (Krissansen-Totton et al. 2021b) and Venus (Krissansen-Totton et al. 2021a). Here, the PACMAN model is modified to explore atmospheric evolution on terrestrial planets around low-mass stars. It is then applied to Trappist-1b, c, d, e, f, and g to make predictions of bulk atmosphere composition consistent with current mass–radius constraints (Agol et al. 2021).

## 2. Methods

The PACMAN model is adapted from Krissansen-Totton et al. (2021b) and is qualitatively summarized in Figure 1. The reader is referred to this paper for the full description of the model. Here, we summarize the key features of the model necessary for interpreting results. The Python code for the model is also available on the lead author's Github upon publication.

Trappist-1 planetary evolution is separated into an initial magma ocean phase (Figure 1, left), and a subsequent solid-mantle phase (Figure 1, right). Post-core formation planets are initialized with a fully molten mantle and some endowment of volatiles, a radionuclide inventory ($^{238}$U, $^{235}$U, $^{232}$Th, $^{40}$K), and an initial mantle oxygen fugacity. The magma ocean freezes from the core, upwards, and hydrogen, carbon, and oxygen are partitioned between dissolved melt phases, crystalline phases, and the atmosphere by assuming chemical equilibrium and melt trapping via compaction (Hier-Majumder & Hirschmann 2017). Heatflow from the interior is calculated using a 1D convective parameterization, with temperature-dependent magma ocean viscosity (see Appendix A).

Planetary oxidation may occur from the net loss of hydrogen to space: hydrogen loss minus half the atomic oxygen escape. Free oxygen produced via net hydrogen escape is dissolved in the melt and may be transferred to the solid mantle as the magma ocean solidifies (Schaefer et al. 2016). We parameterized atmospheric escape as being either diffusion limited or XUV limited, depending on the stellar XUV flux and the composition of the upper atmosphere, which is defined here as everything above the tropopause. During the XUV-driven escape of a steam-dominated atmosphere, the hydrodynamic wind may drag along oxygen (and even $CO_2$), following Odert et al. (2018). In contrast, if the upper atmosphere is mostly dry, then the escape of hydrogen will be limited by the diffusion of hydrogen through the background noncondensable atmosphere, and neither oxygen nor $CO_2$ can escape via thermal processes. Thermosphere temperature and the cold-trap temperature modulate the drag of heavier species and the amount of H-bearing gases that reach upper atmosphere, respectively. We also introduce a nonthermal escape parameterization which allows for up to 100 bar atmospheric erosion over the lifetime of Trappist-1. Escape parameterizations are described in full in Appendix A.

A 1D radiative-convective climate model is used to self-consistently calculate surface temperature, outgoing longwave radiation (OLR), absorbed shortwave radiation (ASR), the water vapor profile, and surface liquid water inventory (if any) during both the magma ocean phase and subsequent temperate evolution. OLR is a function of the surface $H_2O$, $CO_2$, and background $N_2$ inventories, and is calculated using the publicly available correlated-$k$ radiative transfer code of Marcq et al. (2017). Correlated-$k$ coefficients are calculated from the high-resolution molecular absorption spectra computed with *kspectrum* (Eymet et al. 2016), including $H_2O$–$H_2O$ continuum absorption (Clough et al. 2005), and $CO_2$–$CO_2$ continuum absorption (Bézard et al. 2011). To obtain OLR in the presence of condensable water vapor, a dry adiabat to moist adiabat to isothermal atmospheric structure is assumed (Kasting 1988). The temperature of the isothermal upper atmosphere is randomly sampled from $-30$ K to $+30$ K around the planetary skin temperature, which is instellation and orbital separation dependent. This encompasses solar-system variation and is consistent with comprehensive radiative-convective models of Trappist-1 planetary climates (Lincowski et al. 2018).

After a hot start, loss of water to space, secular dimming of the Trappist-1 star, and the dissipation of internal heat from accretion and radionuclides typically cause planetary surface temperatures to drop below the solidus. When this occurs the magma ocean phase is complete, and the model transitions to solid-state mantle convection with temperate geochemical cycling (Figure 1, right). During solid-state evolution, the only source of oxygen remains net hydrogen escape. However, there are now numerous crustal sinks for oxygen including both subaerial and submarine outgassing of reduced species (e.g., $H_2$, CO, $CH_4$), water–rock reactions that generate $H_2$ (when liquid water is present), and dry crustal oxidation. The sizes of these oxygen sinks are self-consistently calculated from the planetary interior evolution and mantle volatile content: outgassing fluxes are calculated using the melt-gas equilibrium outgassing model of Wogan et al. (2020). Outgassing fluxes depend on mantle oxygen fugacity, degassing overburden pressure, the volatile content of the mantle, specifically $H_2O$ and $CO_2$ content, the rate at which melt is produced, and the tectonic regime. Dry and wet crustal sinks for oxygen similarly depend on crustal production rates. To estimate crustal production, a plate tectonics regime is assumed (Krissansen-Totton et al. 2021b). Similar atmosphere–interior models of Venus suggest that the choice of mobile lid and stagnant lid regimes has a comparatively minor effect on atmospheric redox evolution (Krissansen-Totton et al. 2021a). The thermal evolution of the mantle during solid-state evolution is calculated by balancing heat production from radionuclides plus heat flow from the metallic core (imposed) with convective and advective transport to the surface.

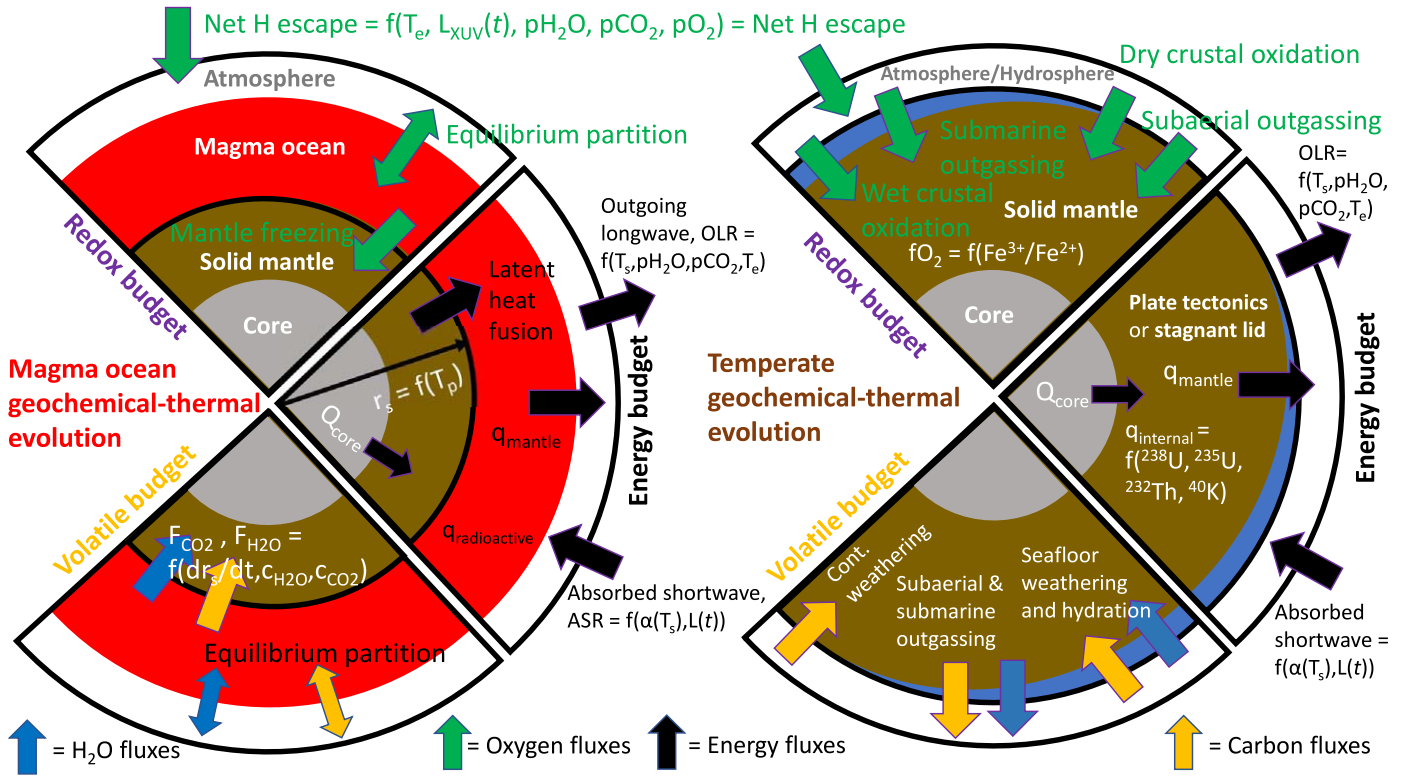

**Figure 1.** Schematic of PACMAN geochemical evolution model applied to Trappist-1 planets. The redox budget, thermal-climate evolution, and volatile budget are modeled from initial magma ocean (left) through to temperate geochemical cycling (right). Oxygen fluxes are shown by green arrows, energy fluxes by black arrows, carbon fluxes by orange arrows, and water fluxes by blue arrows; the net loss of hydrogen to space effectively adds oxygen to the atmosphere. During the magma ocean phase, the radius of solidification, $r_s$, begins at the core–mantle boundary and moves toward the surface as internal heat is dissipated. The rate at which this occurs is controlled by radiogenic and tidal heat production, $q_{internal}$, convective heat flow from the mantle to the surface, $q_{mantle}$, and heat flow from the core, $Q_{core}$. This internal heat flow balances the difference between outgoing longwave radiation (OLR), and incoming absorbed shortwave radiation (ASR). The oxygen fugacity of the mantle, $fO_2$, and the water and carbon content mantle and surface reservoirs are tracked throughout. This schematic is an adapted and updated version from Krissansen-Totton et al. (2021b).

Silicate weathering (Krissansen-Totton et al. 2018a) and the deep hydrological cycle (Schaefer & Sasselov 2015) are also explicitly modeled because climate and surface volatile inventories control crustal oxygen sinks and atmospheric escape processes. Carbon is added to the atmosphere via magmatic outgassing (see above), and—if surface liquid water exists—is returned via continental and seafloor weathering, whose relative contributions depend on climate and the surface water inventory.

### 2.1. Treatment of Volatiles

The model tracks C-, H-, and O-bearing species, as well as $Fe^{3+}/Fe^{2+}$ speciation in the interior. Specifically, surface and mantle inventories of $CO_2$, $H_2O$, and $O_2$ are explicitly calculated. While our outgassing model self-consistently calculates outgassing fluxes of $CO_2$, $H_2O$, $CO$, $H_2$, and $CH_4$ from melt properties (Wogan et al. 2020) and $H_2$ fluxes from serpentinization, all outgassed reductants ($CO$, $H_2$, $CH_4$) are assumed to instantaneously deplete atmospheric oxygen. In the absence of atmospheric $O_2$, all hydrogen from H-bearing reductants is assumed to be rapidly lost to space, therefore net oxidizing the planet; hydrogen escape via outgassed $H_2$ and $CH_4$ will be rapid because, unlike $H_2O$, these species are not cold-trapped. The limitations of our climate module and absence of a photochemical model means we cannot simulate the time evolution of more reducing bulk atmospheres (i.e., $CO$–$H_2$ dominated). An oxidized atmosphere is a common assumption in solar-system terrestrial planet evolution models (Gillmann et al. 2020; Gillmann & Tackley 2014; Zahnle et al. 2010), and polluted white dwarfs suggest similarly oxidizing mantles are ubiquitous (Doyle et al. 2019), but the possible limitations of this assumption are explored in the Section 4. During melt production, volatiles are partitioned between solid and melt phases using a constant partition coefficient for water ($k_{H_2O} = 0.01$), and redox-dependent graphite-saturated partitioning for carbon (see Text G in Krissansen-Totton et al. 2021b, which follows Ortenzi et al. 2020). Nitrogen fluxes are not modeled, and we instead assumed a constant $N_2$ background partial pressure of 1 bar in all nominal model simulations, but sensitivity tests for 0.5–3 bar $N_2$ partial

**Table 1**
Uncertain Parameter Ranges Sampled in Nominal Trappist-1 Monte Carlo Calculations

| | | Nominal range | References/Notes |
|---|---|---|---|
| Initial conditions | Water | $10^{21}$–$10^{23.63}$ kg[a] | 0.7–300 Earth oceans. |
| | Carbon dioxide | $10^{20}$–$10^{22.69}$ kg[a] | Approximately 20 bar–10 kbar, pending other atmospheric constituents and gravity. |
| | Radionuclide U, Th, and K inventory (relative to Earth) | 0.33–30.0[a] | Scalar multiplication of Earth's radionuclide inventories in Lebrun et al. (2013). Allows for modest tidal heating. |
| | Mantle free oxygen | $10^{20.6}$–$10^{22}$ kg[a] | This ensures post-solidification mantle redox within 3–4 log units of the quartz-fayalite-magnetite buffer. |
| Stellar evolution and escape parameters | Trappist-1 XUV saturation time, $t_{sat}$ | $3.14^{+2.22}_{-1.46}$ Gyr | XUV evolution parameters drawn randomly from joint distribution (Birky et al. 2021). |
| | Post-saturation phase XUV decay exponent, $\beta_{decay}$ | $-1.17^{+0.27}_{-0.28}$ | XUV evolution parameters drawn randomly from joint distribution (Birky et al. 2021). |
| | Saturated $\log_{10}(F_{XUV}/F_{BOLOMETRIC})$ flux ratio | $-3.03^{+0.25}_{-0.23}$ | XUV evolution parameters drawn randomly from joint distribution (Birky et al. 2021). |
| | Escape efficiency at low XUV flux, $\varepsilon_{low}$ | 0.01–0.3 | See escape section in Krissansen-Totton et al. (2021b). |
| | Transition parameter for diffusion limited to XUV-limited escape, $\lambda_{tra}$ | $10^{-2}$–$10^{2}$[a] | See escape section in Krissansen-Totton et al. (2021b). |
| | XUV energy that contributes to XUV escape above hydrodynamic threshold, $\zeta_{high}$ | 0%–100% | See escape section in Krissansen-Totton et al. (2021b). |
| | Cold-trap temperature variation, $\Delta T_{cold-trap}$ | −30 to +30 K | Cold-trap temperature, $T_{cold-trap}$, equals planetary skin temperature plus a fixed, uniformly sampled variation, $T_{cold-trap} = T_{eq}(1/2)^{0.25} + \Delta T_{cold-trap}$. Here, $T_{eq}$ is the planetary equilibrium temperature given assumed albedo. |
| | Thermosphere temperature, $T_{thermo}$ | 200–5000 K[a] | (Johnstone et al. 2018, 2021; Lichtenegger et al. 2016). |
| | Nonthermal escape (total loss over Trappist-1 evolution), $NT$ | 1–100 bar[a] | (Dong et al. 2018; Garcia-Sage et al. 2017). |
| Carbon cycle parameters | Temperature-dependence of continental weathering, $T_{efold}$ | 5–30 K | Plausible Earth-like range (Krissansen-Totton et al. 2018a). |
| | $CO_2$-dependence of continental weathering, $\gamma$ | 0.1–0.5 | Plausible Earth-like range (Krissansen-Totton et al. 2018a). |
| | Weathering supply limit, $W_{sup-lim}$ | $10^5$–$10^7$ kg/s[a] | Broad terrestrial planet range (Foley 2015). |
| | Ocean calcium concentration, $[Ca^{2+}]$ | $10^{-4}$–$3 \times 10^{-1}$[a] mol/kg | Plausible range for diverse terrestrial planet compositions (Kite & Ford 2018; Krissansen-Totton et al. 2018a). |
| | Ocean carbonate saturation, $\Omega$ | 1–10 | (Zeebe & Westbroek 2003). |

**Table 1**
(Continued)

| | | Nominal range | References/Notes |
|---|---|---|---|
| Interior evolution parameter | Solid-mantle viscosity coefficient, $V_{coef}$ | $10^1$–$10^3$ Pa s[a] | Solid-mantle kinematic viscosity, $\nu_{rock}$, (m$^2$/s) is given by the following equation: $\nu_{rock} = V_{coef} 3.8 \times 10^7 \exp\left(\frac{350000}{8.314 T_p}\right)/\rho_m$. Here $T_p$ is mantle potential temperature (K) and $\rho_m$ is mantle density (kg/m$^3$). See Krissansen-Totton et al. (2021b). |
| Crustal sinks oxygen and hydrological cycle parameters | Crustal hydration efficiency, $fr_{hydr-frac}$ | $10^{-3}$ to 0.03[a] | Upper limit wt % H$_2$O in oceanic crust. Lower limit hydration limited by cracking. |
| | Dry oxidation efficiency, $f_{dry-oxid}$ | $10^{-4}$ to 10%[a] | Plausible range of processes for Venus (Gillmann et al. 2009) |
| | Wet oxidation efficiency, $f_{wet-oxid}$ | $10^{-3}$ to $10^{-1}$[a] | Based on oxidation of Earth's oceanic crust (Lécuyer & Ricard 1999). |
| | Maximum fractional molten area, $f_{lava}$ | $10^{-4}$ to 1.0[a] | See explanation in Krissansen-Totton et al. (2021b). |
| | Max mantle water content, $M_{solid-H_2O-max}$ | 0.5–15 Earth oceans | Best estimates maximum hydration of silicate mantle (Cowan & Abbot 2014). |
| Albedo parameters | Hot state albedo (during runaway greenhouse/magma ocean), $A_H$ | 0.0–0.2 | (Pluriel et al. 2019). |
| | Cold state albedo (during temperate state), $A_C$ | 0.0–0.5 | (Kopparapu et al. 2017; Macdonald et al. 2022; Rushby et al. 2020; Shields et al. 2013). |

**Note.**
[a] Denotes this variable was sampled uniformly in log space. All others (except stellar XUV parameters) were sampled uniformly in linear space.

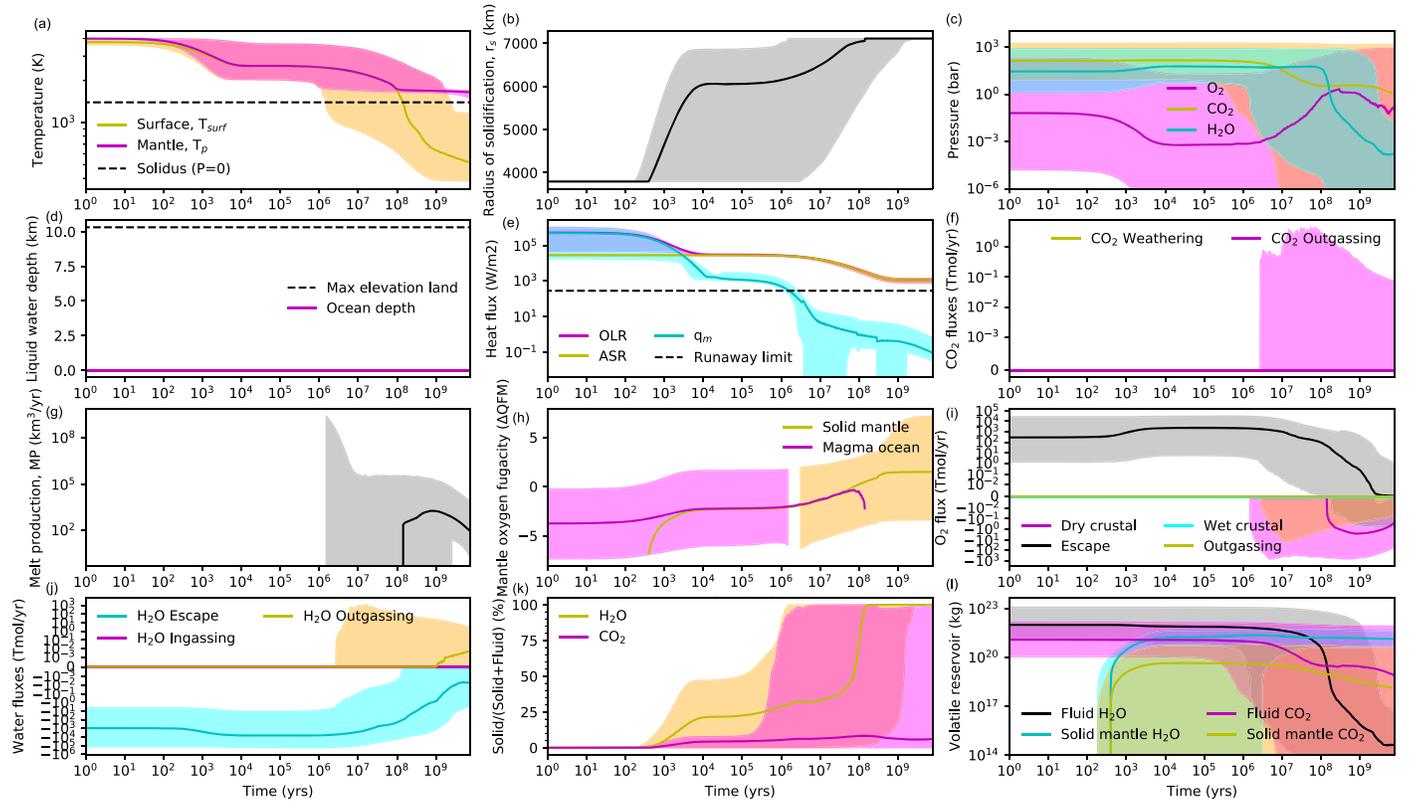

**Figure 2.** Trappist-1b predicted atmosphere–interior time evolution. The lines are median values and shaded regions denote 95% confidence intervals. The time evolution spans from post-accretionary magma ocean to the present day ($8 \times 10^9$ yr). Subplot (a) shows the evolution of mantle potential temperature (magenta) and surface temperature (orange) alongside the silicate solidus (black dashed line). The magma ocean (b) persists from anywhere between a few million years to a few billion years, depending primarily on initial water endowment. Water catastrophically degasses from the magma ocean and is subsequently lost to space via hydrodynamic escape (c, j). Trappist-1b's small stellar separation means the absorbed shortwave radiation never drops below the runaway greenhouse limit (e), and so liquid water never condenses on the surface (d). Atmospheric oxygen may be produced via escape (c, i), but this oxygen is drawn down by geological sinks (i), namely magmatic outgassing of reduced species (orange) and dry crustal oxidation (magenta). The silicate mantle is gradually oxidized by this net hydrogen loss (h). Volatile cycling is controlled by the rate at which fresh crust is produced (g), and while melt production typically continues throughout Trappist-1b's evolution, outgassing fluxes are usually small because the mantle is left comparatively desiccated after the magma ocean (l).

pressures are also presented. Sulfur species are ignored in our model because, for Earth-like bulk abundances, outgassing of S-bearing gases only modestly increases total oxygen sinks (Krissansen-Totton et al. 2021b).

### 2.2. Monte Carlo Calculations and Unknown Parameters

There are many unknown parameters and initial conditions in the model, including temperature-dependent mantle viscosities, efficiencies of XUV-driven escape, uncertain stellar XUV fluxes, carbon cycle feedbacks, deep hydrological cycle dependencies, and albedo parameterizations. We adopted a Monte Carlo approach where we ran the model thousands of times uniformly sampling all 26 unknown parameters. The parameter ranges for key variables are shown in Table 1, and important variables are described below.

Initial volatile inventories are a key control on Trappist-1 planetary evolution. For each planet modeled, we assumed an initial water inventory anywhere from 0.7 to 300 Earth oceans (uniformly sampled in log space). We similarly assume planetary $CO_2$ endowments anywhere from 20 bar–10,000 bar, and initial free oxygen sampled from $10^{20.6}$–$10^{22}$ (kg), which ensures an evolved mantle redox state around the quartz-fayalite-magnetite buffer, but with an approximately 8 log unit spread.

The efficiencies of crustal oxygen sinks are controlled by several parameters. Dry crustal oxidation efficiency, $f_{\mathrm{dry-oxid}}$, is the fraction of $Fe^{2+}$ in newly produced crust that is directly oxidized to $Fe^{3+}$ in the presence of an oxidizing atmosphere via nonaqueous reactions. This parameter is sampled uniformly in log space from $10^{-4}$ to 10%, which accommodates a range of physical processes including the diffusion of oxygen into extrusive lava flows (Gillmann et al. 2009), oxidation of small-grain erosion products (Arvidson et al. 1992), and other gas-solid redox reactions (Zolotov 2019), including aerosol oxidation following explosive volcanism (Warren & Kite 2021). The efficiency of crustal oxidation via water–rock reactions, $f_{\mathrm{wet-oxid}}$ (sampled from $10^{-3}$ to $10^{-1}$), represents the fraction of crustal iron that is oxidized via serpentinizing hydration reactions (Lécuyer & Ricard 1999). The magnitude of these crustal sinks, as well as the size of magmatic outgassing sinks, are ultimately limited by rates of crustal production, which are controlled by internal heating. Here, we assume initial radionuclide endowments anywhere from one-third to 30 times that of the Earth. This two order of magnitude range is intended to accommodate possible tidal heating contributions, which are not explicitly modeled (See Section 4).

Instead of explicitly modeling cloud and haze feedbacks, which is a complex 3D problem (Fauchez et al. 2020a; Turbet et al. 2018; Wolf 2017), we assume a constant albedo for both runaway greenhouse (ranging from 0 to 0.2) and temperate states (ranging from 0 to 0.5) with a smooth transition between the two (Pluriel et al. 2019). This constant albedo assumption neglects feedbacks between bulk atmospheric composition and planetary reflectivity, but it is adequate for sampling the broad range of climate states attributable to varying aerosol properties and surface coverage (Shields et al. 2013; Kopparapu et al. 2017; Rushby et al. 2020; Macdonald et al. 2022).

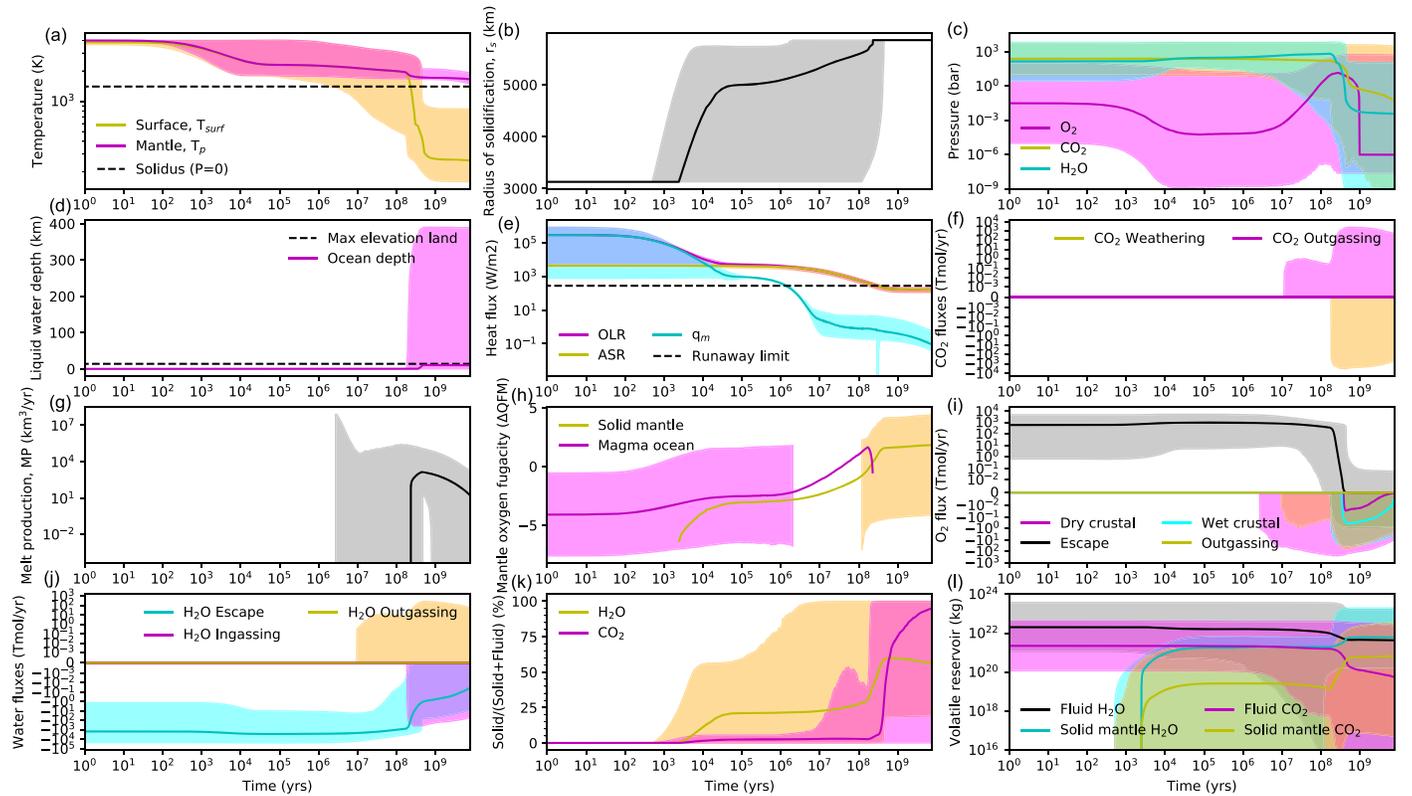

**Figure 3.** Trappist-1e predicted atmosphere–interior time evolution. The lines are median values and shaded regions denote 95% confidence intervals. The time evolution spans from post-accretionary magma ocean to the present day ($8 \times 10^9$ yr). Subplot (a) shows the evolution of mantle potential temperature (magenta) and surface temperature (orange) alongside the silicate solidus (black dashed line). The magma ocean (b) persists from anywhere between a few million years to a few hundred million years, depending primarily on initial water endowment and albedo. Water catastrophically degasses from the magma ocean and is subsequently lost to space via hydrodynamic escape (c, j). The runaway greenhouse phase ends when Trappist-1 dims sufficiently for absorbed shortwave radiation to drop below the runaway greenhouse limit (e). When this occurs, atmospheric water vapor (c) condenses onto the surface producing an ocean (d). A temperate carbon cycle also commences (f), which sequesters carbon in the mantle (k), and volatile cycling is controlled by the rate at which fresh crust is produced (g). Once the steam atmosphere has condensed, atmospheric oxygen produced via escape during the pre-main sequence (i, j) may be drawn down by geological sinks (i): magmatic outgassing of reduced species (orange), water–rock reactions (cyan), and dry crustal oxidation (magenta). In most cases, these sinks are sufficient to remove virtually all oxygen from the atmosphere (c).

The time evolution of the bolometric luminosity of Trappist-1 is adopted from Baraffe et al. (2015), whereas the XUV evolution is adopted from Birky et al. (2021). Specifically, we randomly sample the stellar parameter joint distribution derived in Birky et al. (2021), meaning Trappist-1 has an XUV saturation time of $3.14^{+2.22}_{-1.46}$ Gyr, an XUV-to-bolometric ratio of $-3.03^{+0.25}_{-0.23}$ (in $\log_{10}$ space) during the saturation phase, and a XUV decay exponent of $-1.17^{+0.27}_{-0.28}$. Figure A1 shows the assumed bolometric and XUV luminosity evolution of Trappist-1. For ease of final state comparisons, a precise age of 7.96 Gyr is assumed since the other stellar parameter distributions are largely independent of age, and because sensitivity tests in Appendix D further confirm assumed stellar age has a minimal impact on atmospheric outcomes. Atmospheric thermal escape parameters are discussed in Appendix A. To represent nonthermal escape, we assumed a constant rate of loss totaling 1–100 bar over the 8 Gyr lifetime of Trappist-1 (sampled log-uniformly), where atmospheric constituents are not fractionated, for simplicity, i.e., nonthermal losses of $CO_2$, $O_2$, and $H_2O$ are weighted by mixing ratio.

Monte Carlo calculations were performed for Trappist-1b, c, d, e, f, and g (h is excluded since the PACMAN model does not yet incorporate multiple condensible species). For each planet, 5000 model runs were calculated, and only those that satisfied observed mass–radius constraints were retained. Specifically, we adopted the mass constraints and inferred maximum surface volatile mass fractions from Agol et al. (2021). For the inner planets (b, c, and d), maximum water mass fractions are small: <0.13%, <0.08%, and <0.004%, respectively, given a metallic core mass fraction of up to 50% (Agol et al. 2021). This is because inside the runaway greenhouse limit any surface water is present as steam, which would inflate planetary radius for large water inventories (Turbet et al. 2020b). For the outer planets (e, f, and g), surface volatile mass fractions could be much larger (<11.6%, <14.0%, <16.0%, respectively), for a core mass fraction up to 50%, due to the possibility of liquid or frozen surface water. Note that we assumed that this mass fraction constraint applies to the sum of all surface volatiles, not merely water. Additionally, while we are allowing for high maximum metallic core fractions compared to non-Mercury solar-system bodies (<50%) in our nominal calculations, we also implemented sensitivity tests for more Earth-like metallic cores. Specifically, for core mass fractions <32.5%, surface volatile mass fractions are <0.001% for all the inner planets, and <4.6%, <6.3%, and <8.4% for e, f, and g, respectively. Note that for 1d the small surface volatile mass fraction only applies to atmospheric volatiles since it is possible that 1d retains a liquid water ocean if albedo is high.

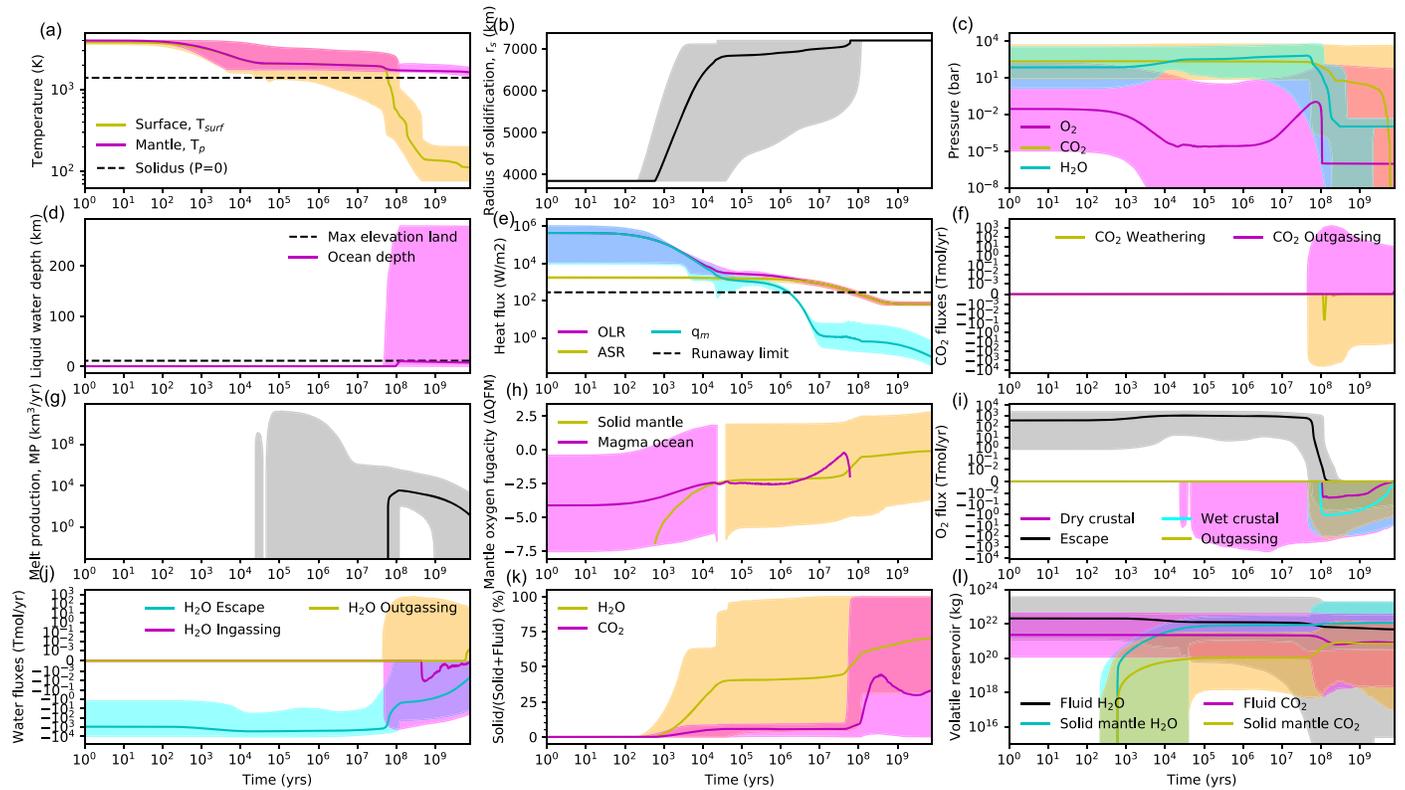

**Figure 4.** Trappist-1g predicted atmosphere–interior time evolution. The lines are median values and shaded regions denote 95% confidence intervals. The time evolution spans from post-accretionary magma ocean to the present day ($8 \times 10^9$ yr). Subplot (a) shows the evolution of mantle potential temperature (magenta) and surface temperature (orange) alongside the silicate solidus (black dashed line). The magma ocean (b) persists from anywhere between a few tens of thousands of years to 100 million years, depending primarily on initial water endowment and Bond albedo. Water catastrophically degasses from the magma ocean and is subsequently lost to space via hydrodynamic escape (c, j). The runaway greenhouse phase ends when Trappist-1 dims sufficiently for absorbed shortwave radiation to drop below the runaway greenhouse limit (e). When this occurs, any remaining atmospheric water vapor (c) condenses onto the surface producing an ocean or frozen crust (d). A temperate carbon cycle also commences (f, k), where volatile cycling is controlled by the rate at which fresh crust is produced (g). Once the steam atmosphere has condensed, atmospheric oxygen produced via escape during the pre-main sequence (c, i) may be drawn down by geological sinks (i): magmatic outgassing of reduced species (orange), water–rock reactions (cyan), and dry crustal oxidation (magenta). In most cases, these sinks are sufficient to remove virtually all oxygen from the atmosphere (c).

## 3. Results

Figure 2 shows the atmosphere–interior evolution of Trappist-1b, as calculated using the PACMAN model. Only model runs that satisfy mass–radius constraints (volatile mass fraction <0.13%) are shown. The high instellation received by Trappist-1b leads to a magma ocean that persists for anywhere from a few million years to several billion years (Figure 2(b)), depending on water inventory and the efficiency of atmospheric escape. The median duration of the magma ocean is around 100 Myr before enough water is lost for the surface to crystallize. Liquid water never condenses due to high instellation (Figures 2(d), (e)), and so there can be no silicate weathering (Figure 2(f)) or water–rock reactions (Figure 2(i)). However, the gradual escape of hydrogen to space (Figures 2(i) and (j)) net oxidizes the planet over geologic time; this may or may not leave behind an $O_2$-rich atmosphere (Figure 2(c)) depending on the efficiency of sinks and initial water inventory (see below). The net oxidation of the mantle due to hydrogen escape and crustal sinks for oxygen is shown in the evolution of mantle redox state (Figure 2(h)). Atmospheric $CO_2$ also degases from the magma ocean and may be subsequently eroded by XUV drag and nonthermal escape; there is a wide spread in modern abundances (Figure 2(c)), depending largely on initial $CO_2$ endowments (see below).

Figure 3 shows the atmosphere–interior evolution of Trappist-1e. Since Trappist-1e resides within the main-sequence habitable zone of Trappist-1, the magma ocean lasts ~1 Myr to a few hundred million years, at most (Figure 3(b)). Once the instellation received drops below the runaway greenhouse limit (Figure 3(e)), any water remaining in the atmosphere condenses out to form a liquid water ocean at the surface (Figures 3(c), (d)). This enables a temperate carbon cycle to regulate atmospheric $CO_2$ (Figure 3(f)), as well as water–rock reactions to draw down oxygen (Figure 3(i)). Once the steam atmosphere has collapsed, hydrogen escape is dramatically lowered due to water being cold-trapped in the troposphere; this means less oxygen production after magma ocean crystallization and so crustal sinks of oxygen dominate (Figure 3(i)). Consequently, anoxic modern atmospheres are the most probable outcome (Figure 3(c)). With that said, there are many model runs where oxygen is left over from the early steam until the present; this is broadly consistent with the magma ocean modeling in Barth et al. (2021). Note that uncertainties in surface volatile inventories for 1e are larger

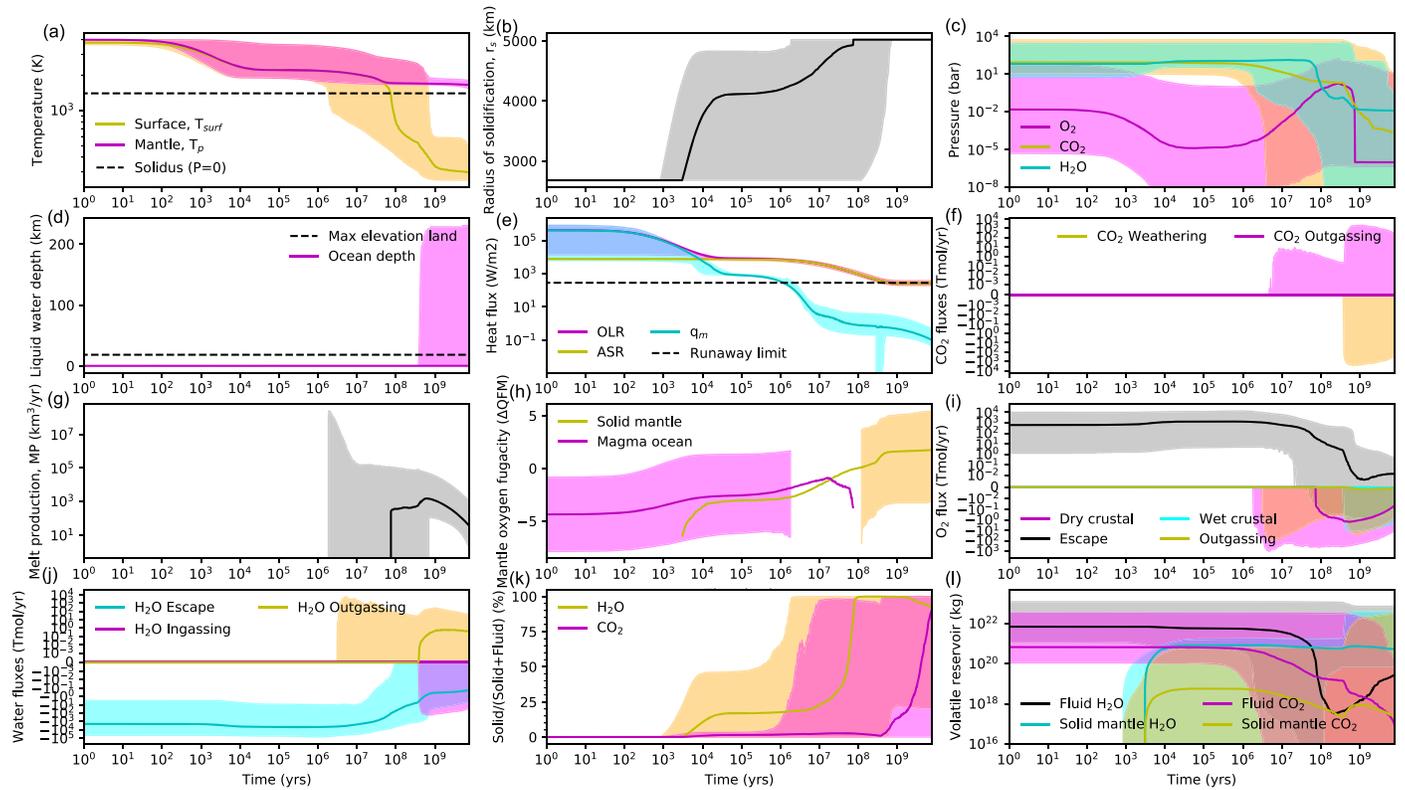

**Figure 5.** Trappist-1d predicted atmosphere–interior time evolution. The lines are median values and shaded regions denote 95% confidence intervals. The time evolution spans from post-accretionary magma ocean to the present day ($8 \times 10^9$ yr). Subplot (a) shows the evolution of mantle potential temperature (magenta) and surface temperature (orange) alongside the silicate solidus (black dashed line). The magma ocean (b) persists from anywhere between a few million years to a few hundred million years, depending primarily on initial water endowment and Bond albedo. Water catastrophically degasses from the magma ocean and is subsequently lost to space via hydrodynamic escape (c, j). The runaway greenhouse phase may end when Trappist-1 dims sufficiently for absorbed shortwave radiation to drop below the runaway greenhouse limit (e). However, this only occurs in around half of models runs—if albedo is low, then a runaway greenhouse persists until virtually all water is lost to space. If the runaway greenhouse threshold is crossed, any remaining atmospheric water vapor condenses onto the surface producing an ocean (d). A temperate carbon cycle also commences (f), which sequesters carbon in the mantle (k), and volatile cycling is controlled by the rate at which fresh crust is produced (g). If the steam atmosphere condenses, atmospheric oxygen produced via escape during the pre-main sequence (i, j) may be drawn down by geological sinks (i): magmatic outgassing of reduced species (orange), water–rock reactions (cyan), and dry crustal oxidation (magenta). In most cases, these sinks are sufficient to remove virtually all oxygen from the atmosphere (c).

than those of 1b, since the water mass fraction of 1b is better constrained (see methods).

Figure 4 shows the atmosphere–interior evolution of Trappist-1g. The results are similar to 1e, except that, because the orbital separation is greater, the average magma ocean duration is even shorter (Figure 4(b)), and 1g is colder post-magma ocean (Figure 4(a)). In fact, in most model runs a frozen surface is likely, leading to continuous $CO_2$ drawdown via seafloor weathering, and an even colder surface climate. Cold climates enhance the water cold trap, implying negligible oxygen production via hydrogen escape post-magma ocean crystallization (Figure 4(i)). Consequently, anoxic modern atmospheres are highly probable (Figure 4(c)), although $O_2$-rich atmospheres occasionally persist due to early accumulation and limited subsequent sinks.

Similar calculations were repeated for 1c and 1f, and these are shown in Appendix B. In summary, the predicted evolution of 1c is very similar to 1b, whereas 1f is similar to 1g.

Trappist-1d is an unusual intermediate case because of its small size, and because it straddles the installation threshold where high albedo states could conceivably produce temperate climates. Figure 5 shows the atmospheric evolution of Trappist-1d, as calculated using the PACMAN model. Only model runs that satisfy mass–radius constraints (atmospheric mass fraction <0.004%) are shown. Model outputs are a combination of 1c-like outputs and 1e-like outputs, since the orbital separation of 1d is such that approximately half of all model runs are in perpetual runaway greenhouse and half are more temperate. Note that because albedo is an arbitrary Monte Carlo parameter in our calculations, these results do not directly inform the habitability prospects of planet 1d.

### 3.1. Evolved Atmosphere Predictions

Figure 6 shows probability distributions for the modern compositions of the Trappist-1b–g atmospheres. In other words, Figure 6 shows the spread of atmospheric bulk abundances in Figures 2(c), 3(c), 4(c), and 5(c) at the end of model evolution. As noted previously, these are testable predictions because dense $O_2$, $CO_2$, or steam-rich atmospheres may be detectable with JWST (Gialluca et al. 2021; Krissansen-Totton et al. 2018b; Lincowski et al. 2018; Lustig-Yaeger et al. 2019a; Morley et al. 2017; Wunderlich et al. 2019, 2020). Figure 6 quantifies the broad trends described above: abiotic $O_2$ buildup is most probable for 1b and 1c, although a comparable number of model runs also result in anoxic atmospheres. There is a bimodal distribution of $CO_2$ for 1b and 1c, which is largely controlled by initial inventory and atmospheric escape assumptions (see below). The atmospheric water vapor distribution is also bimodal:

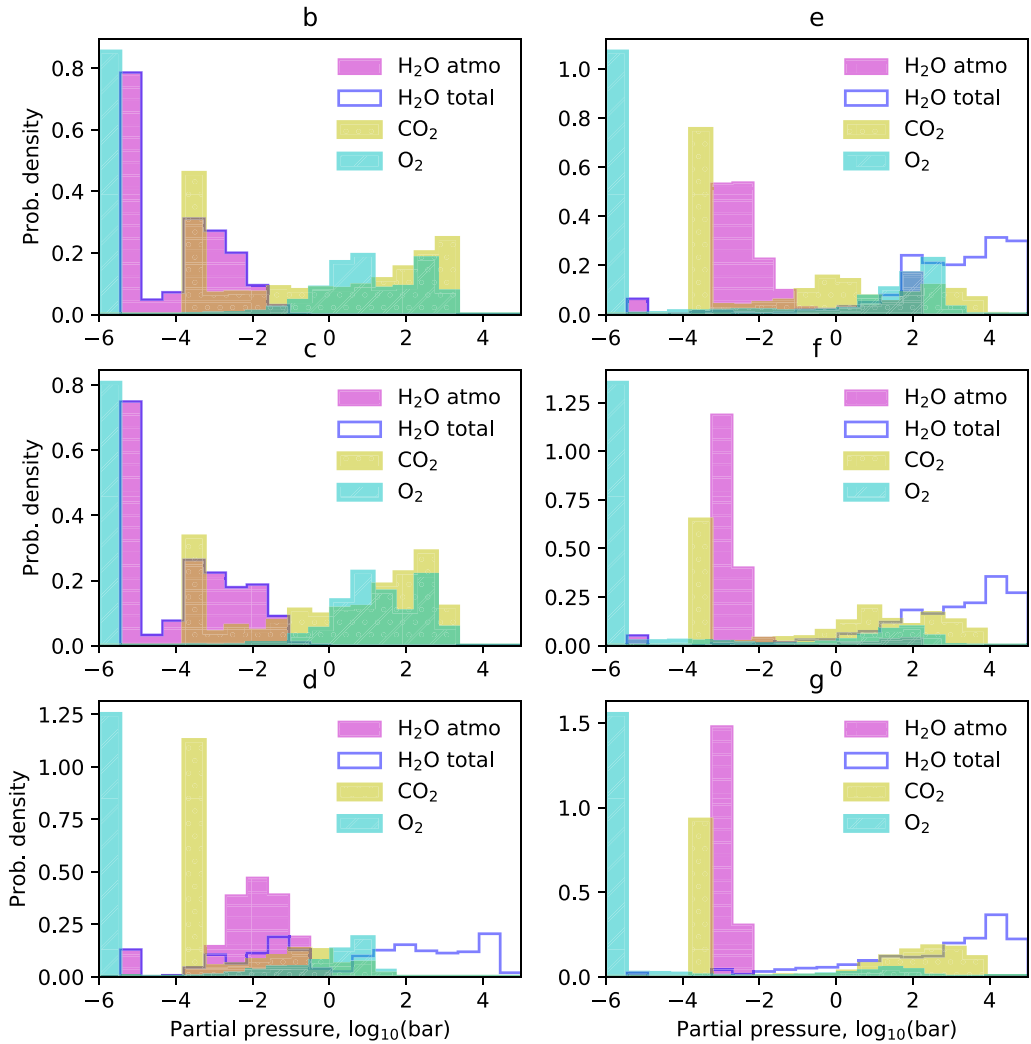

**Figure 6.** Probability distributions for modern atmospheric abundances of Trappist-1 planets as predicted by the PACMAN geochemical evolution model. Partial pressure distributions for carbon dioxide (yellow), oxygen (cyan), and water vapor (magenta) are shown for 1b, c, d, e, f, and g. The total surface water inventory is denoted by the unfilled, blue line distribution; this is identical to the water vapor distribution for the inner planets, but not for the outer planets because surface water can condense as a liquid or solid. Note that there are numerical cutoffs for carbon dioxide, oxygen, and water vapor at $10^{-4}$, $10^{-6}$, and $10^{-6}$, respectively. Broadly speaking, the probability of oxidized atmospheres decreases with planet–star separation.

steam is either entirely lost to space or persists at some low (<0.01 bar) partial pressure due to escape being throttled by diffusion through background $CO_2$ or $O_2$. Note that atmospheric $H_2O$ equals total surface $H_2O$ for 1b and 1c because liquid water cannot condense at the surface.

For 1d, the planet's small size limits the maximum gaseous envelope even more severely than 1b and c (when all surface volatiles are gaseous). Additionally, if high albedo states are possible, then surface liquid water condensation can occur, so the distributions in 1d are a mixture of temperate climates and runaway greenhouse states. Generally speaking, however, the combination of lower total instellation, less XUV received, and shorter-duration magma oceans leads to less abiotic $O_2$ buildup compared to 1b and c, and lower initial water inventories from mass–radius constraints only compound this.

For the Trappist-1 planets that are unambiguously in the main-sequence habitable zone (1e, f, and g), anoxic atmospheres are the most likely atmospheric outcome, and these become increasingly probable with greater stellar separation (Figure 6). This trend arises because the time taken for absorbed shortwave radiation to fall below the runaway greenhouse decreases with increasing stellar separation; a shorter-duration magma ocean means less hydrogen escape, and the presence of liquid surface water means potentially larger total oxygen sinks via water–rock reactions. For the outer planets, the biomodal distribution in atmospheric $CO_2$ is attributable to scenarios where $CO_2$ drawdown is continuous and scenarios where a silicate weathering thermostat maintains a temperate climate. Virtually no atmospheric water vapor is anticipated for 1f and g due their cold climates, but large surface reservoirs of water (either liquid or frozen) are permissible.

### 3.2. Key Parameters Governing Divergent Evolved Atmospheres

Given the complexity of the PACMAN model and the large number of uncertain, governing parameters (Table 1), understanding the causal relationships behind different evolved atmospheres is challenging. Figure 7 shows some example scatter plots (with associated linear regressions) illustrating relationships between final atmospheric $O_2$ after 8 Gyr and initial water (top), dry oxidation efficiency (middle), and albedo (lower) for planets 1b (left) and e (right). Several insights emerge from these figures. First, there is strong relationship between initial water inventory and the likelihood of abiotic oxygenation. This is because the more water there is to lose during the early period of high XUV fluxes, the more $O_2$ is potentially left behind to be mopped up by crustal sinks. Additionally, more atmospheric water vapor means a longer-duration magma ocean, and thus a more persistent steam atmosphere and greater cumulative hydrogen escape. Second, there is a negative correlation between dry oxidation efficiency, $f_{dry-oxid}$, and initial water. These figures show the value of a simple linear correlation because the relationship between oxygen and dry oxidation efficiency for both 1b and e is highly statistically significant ($p \sim 10^{-50}$ and $p \sim 10^{-6}$, respectively) even though the trend is not immediately obvious by inspection. Dry oxidation efficiency is a strong predictor of atmospheric oxygenation for inner planets such as 1b because there are no water–rock sinks given the lack of liquid surface water. Third, causal explanations for atmospheric oxygenation are different for different Trappist-1 planets. In particular, albedo matters for 1e (highly significant negative correlation) because albedo dictates the duration of the magma ocean and runaway greenhouse, but not for 1b, which is always in runaway greenhouse regardless of albedo, and so there is no correlation between albedo at atmospheric oxygenation.

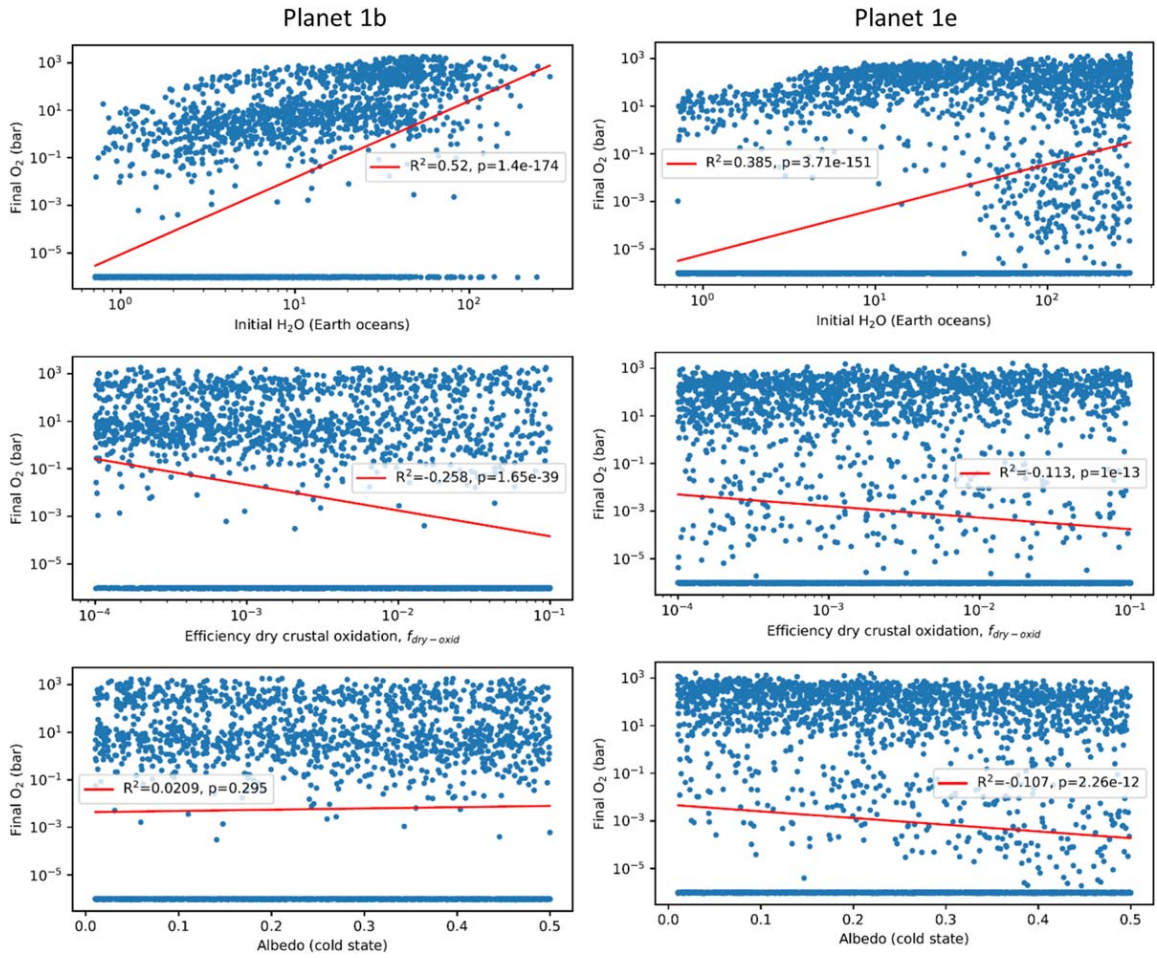

**Figure 7.** Correlation between atmospheric oxygen after 8 Gyr and three input parameters: initial water inventory (top), dry crustal oxidation efficiency (middle), and temperate state albedo (below). Left column denotes Trappist-1b and right column 1e. Each blue dot is an individual model run, and red lines show results from a linear correlation, with corresponding $R^2$ values and $p$-values shown as legends. Note the numerical cutoff at $pO_2 = 10^{-6}$ bar for computational efficiency.

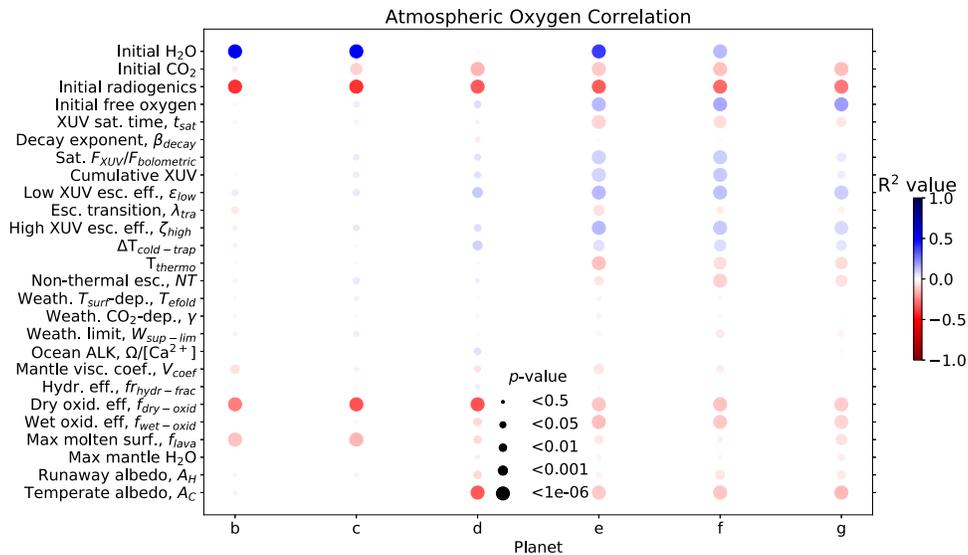

**Figure 8.** Linear correlations between atmospheric oxygen after 8 Gyr of evolution and the input parameters from Table 1 (rows). Columns denote results for Trappist planets 1b, c, d, e, f, and g. The color of each dot shows the $R^2$ value of the linear correlation, whereas the size of the dot reflects the statistical significance of the relationship; larger dots imply a more significant linear relationship. The redox evolution of the inner planets 1b and c is controlled almost completely by initial volatile inventories, the efficiency of dry crustal oxidation, and interior heat production. In contrast, the redox evolution of the outer planets is more complex; it depends on the physics of atmospheric escape, albedo feedbacks, initial redox state, in addition to initial volatile inventories and sink efficiencies.

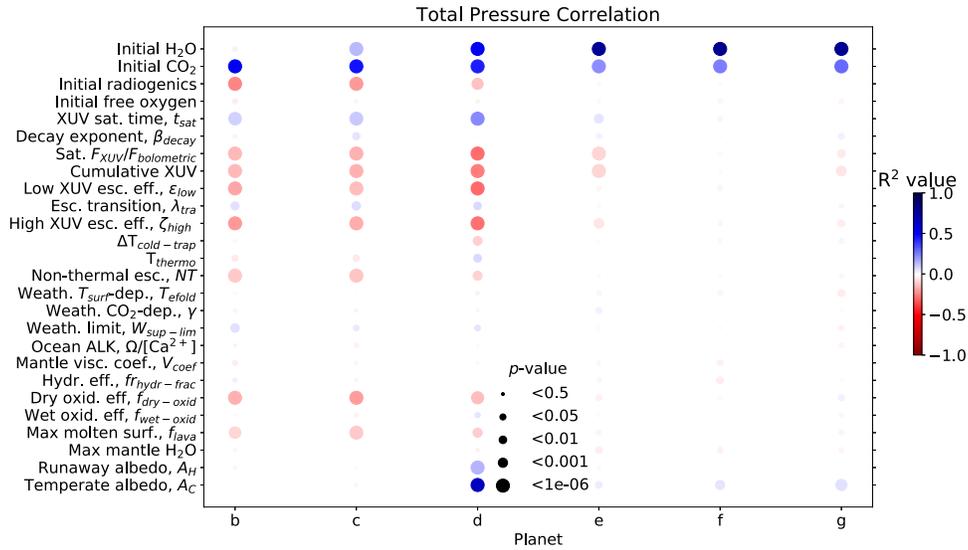

**Figure 9.** Linear correlations between total surface volatile inventories after 8 Gyr of evolution and the input parameters from Table 1 (rows). Columns denote results for Trappist-1 planets 1b, c, d, e, f, and g. The color of each dot shows the $R^2$ value of the linear correlation, whereas the size of the dot reflects the statistical significance of the relationship (larger dots imply a more significant linear relationship). Atmospheric retention on the outer planets is controlled almost entirely by initial volatile inventories. However, for the inner planets, the details of atmospheric escape, the efficiency of crustal oxygen sinks, and internal heat production all influence the retention of substantial secondary atmospheres.

Instead of laboriously presenting dozens of scatter plots like Figure 7, a correlation plot can be used to understand patterns of atmospheric oxygenation in the Trappist-1 system. Figure 8 shows linear correlations between all the unknown input parameters in Table 1 and final atmospheric oxygen abundance for the six Trappist-1 planets considered here. The color of each dot denotes the $R^2$ value for that correlation, which reflects both the sign and strength of that relationship. The sizes of the dots are related to the statistical significance of the trend: the larger the dot, the more statistically significant the relationship between two variables. Note that choosing different lower numerical cutoffs, such as $10^{-6}$ or $10^{-10}$ for oxygen, has a negligible effect on such correlation plots (not shown).

Several interesting patterns emerge from Figure 8. Trappist-1b and c are seemingly simple objects. All that matters for the purposes of anticipating atmospheric oxygenation is initial water inventory, dry oxidation efficiency, and internal heating (via radionuclides or tidal heating). These relationships are straightforward to understand: more initial water implies greater potential hydrogen loss, greater dry oxidation efficiency means melt production causes greater oxygen drawdown, whereas more internal heat production leads to more melt production post-magma ocean, and therefore greater outgassing and dry crustal oxygen sinks. The maximum surface magma fraction, $f_{lava}$, which controls dry oxidation immediately after magma ocean solidification when melt production is large, is also negatively correlated with oxygenation.

In contrast, abiotic oxygen accumulation is much more complex to predict for the outer planets because there are a dozen or so important parameters correlated with oxygen. The positive correlation with initial water inventory largely remains. But atmospheric oxygen is additionally correlated with escape parameters that control the oxygen source: the efficiency of XUV-driven escape, $\varepsilon_{low}$, is positively correlated with atmospheric $O_2$, as is the fraction of XUV energy above the O-drag threshold that goes into driving addition escape as opposed to being reradiated to space or conducted to the lower atmosphere, $\zeta_{high}$. There are also correlations with other atmospheric parameters governing escape, such as thermosphere temperature (negatively correlated), the cumulative XUV flux (positively correlated), and total nonthermal escape (negatively correlated). Similar to Figure 7(f), oxygen accumulation on all outer planets is negatively correlated with albedo for the reasons explained above. Oxygen accumulation is negatively correlated with initial $CO_2$ for the outer planets because more atmospheric $CO_2$ results in more efficient cold-trapping of water and thus less hydrogen escape. Interestingly, initial mantle redox is positively correlated with atmospheric oxygen for the outer planets, but not for the inner planets. This is likely because the inner planets are left desiccated by their long-duration, highly irradiated magma ocean phases. Subsequent outgassing sinks are low because the mantle retains few volatiles (compare Figure 2(i) with Figures 3(i) and 4(i)), and so mantle redox state matters less for the size of crustal oxygen sinks after magma ocean crystallization. Also on the sink side, Figure 8 shows that while dry oxidation efficiency matters somewhat for the outer planet oxygenation, the efficiency of oxygen consumption via water–rock reactions is more strongly inversely correlated with atmospheric oxygen. This is consistent with the complete submergence of all land on about half of outer planet model runs (Figures 3(d), 4(d)). In summary, explaining patterns of oxygenation on the habitable zone terrestrial planets is somewhat more complicated, but it should be remembered that the overall likelihood of oxygen buildup on these worlds is comparatively low (Figure 6).

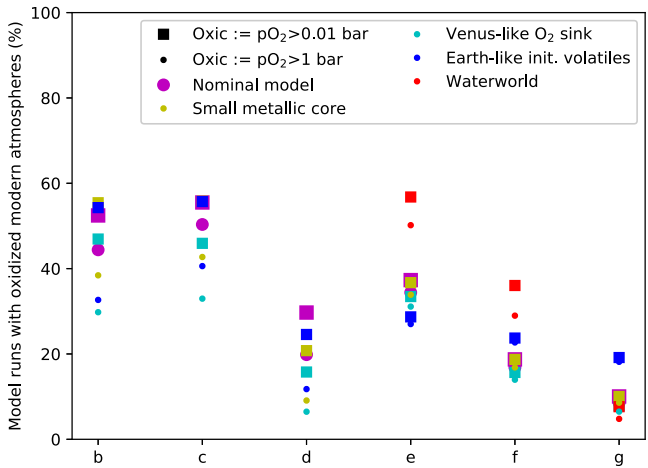

**Figure 10.** Percentage of model runs with oxygen-rich atmospheres after 8 Gyr of evolution for planets 1b–g. Inner planets 1b and c are left with oxygen-rich atmospheres in about half of all model runs, whereas 1d–g are typically anoxic after 8 Gyr. Squares denote results when "oxygen-rich" is defined as $pO_2$ exceeding 0.01 bar, whereas circles denote a threshold of 1 bar. The latter is the approximate detectability threshold for JWST (Fauchez et al. 2020b). Different colors represent model runs from nominal calculations (magenta), the small metallic core sensitivity test (yellow), the Venus-like efficient dry oxidation sink sensitivity test (cyan), the Earth-like initial volatile inventory test (blue), and waterworld outer planets (red).

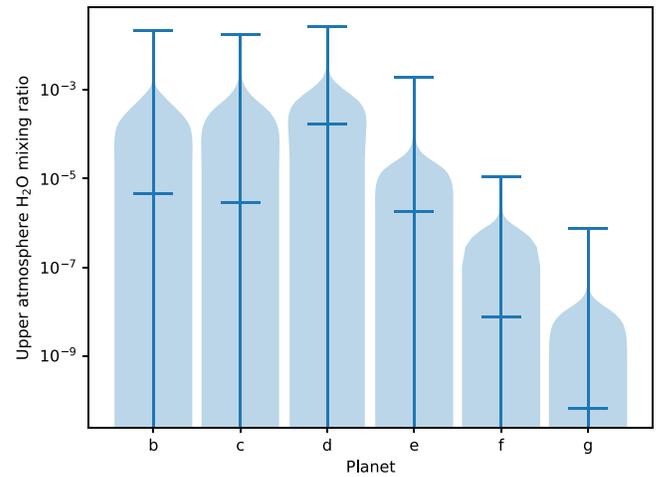

**Figure 11.** Violin plot probability distributions of upper-atmosphere water vapor for planets 1b–g from nominal model runs. Horizontal blue lines denote median and extrema values. Atmospheric water vapor is typically only potentially detectable in transit spectra for 1b, c, and d.

Figure 9 shows a similar correlation plot, except this time the linear calculation between each unknown variable and the total surface volatile inventory has been calculated. This is a crude measure of what variables govern whether or not each planet is likely to retain a substantial secondary atmosphere. Unsurprisingly, for virtually every planet, there is a strong positive correlation with initial water and $CO_2$ endowments. Additionally, for the inner planets, atmospheric retention is strongly dependent on the variables that control thermal and nonthermal escape, whereas for the outer planets, atmospheric retention is largely independent of the details of escape. This difference can be attributed to the lower XUV fluxes received by outer planets and the finite duration of their steam-dominated runaway greenhouse atmospheres. Finally, for the inner planets, the efficiency of oxygen sinks (e.g., $f_{dry-oxid}$ and $f_{lava}$) also influences the total atmospheric mass after 8 Gyr of evolution.

Similar correlation plots can be produced to understand patterns of atmospheric $CO_2$ abundances and atmospheric water vapor (Appendix C). Here, final atmospheric $CO_2$ is governed by initial inventory (all planets), thermal escape physics (inner planets only), and carbon cycle feedbacks (outer planets only).

### 3.3. Sensitivity Tests

The atmospheric evolution calculations described above were repeated with smaller assumed metallic core fractions, more akin to solar-system terrestrials (<32.5% core mass fractions). Smaller core fractions imply smaller maximum surface volatile inventories, especially for the inner planets where liquid water cannot condense on the surface. In fact, iron core fractions <32.5% imply atmospheric volatile mass fractions for 1b, c, and d that are <0.001% by mass. When this constraint is applied, the distributions for final atmospheric abundances (Figure D1 in Appendix D) are qualitatively similar to Figure 6. Dense, $CO_2$-rich atmospheres are less probable for the inner planets assuming smaller metallic cores, but the correlations with atmospheric oxygen are similar (Figure D2), and implied initial volatile inventories also do not change substantially compared to the nominal model (compare Figures A3 and D3).

For the outer planets, we consider an additional sensitivity test where the comparatively low densities of the outer planets (relative to Earth) are assumed to be attributable to large surface volatile inventories rather than a smaller core mass fraction. Mass–radius constraints imply core mass fractions exceeding 25% would require large surface water inventories for all three outer planets, and for 1f, even an 18% core mass fraction requires large surface volatile fractions (Agol et al. 2021). For this waterworlds sensitivity test we adopt a 25% core mass fraction minimum for all three outer planets (and 50% maximum), which implies modern water mass fractions satisfying $1.1\% < wf < 11.6\%$ for 1e, $2.4\% < wf < 14\%$ for 1f, and $0.57\% < wf < 16\%$ for 1g (Agol et al. 2021). Time evolution plots are shown for all three planets in Appendix D. Consistent with previous work, we find large surface volatile inventories suppress oxygen sinks such as volcanism, both via shutting down melt production (Kite & Ford 2018; Krissansen-Totton et al. 2021b) and through limiting exsolution of reducing volatiles from partial melts (Krissansen-Totton et al. 2021c).

Consequently, after initial $O_2$ accumulation during the pre-main sequence runaway greenhouse phase, it becomes more difficult to remove atmospheric oxygen subsequently, and so oxygen-rich modern atmospheres are somewhat more likely for the outer planets if they are waterworlds.

When the PACMAN model was applied to Venus's evolution, it was found that comparatively efficient dry crustal efficiency ($f_{dry-oxid} > 10^{-3}$) is required to recover the modern, oxygen-free Venus atmosphere (Krissansen-Totton et al. 2021a). If this constraint is applied to the Trappist-1 planets, then oxygen accumulation becomes slightly less probable for all planets (Figure D7), but the qualitative patterns of oxygenation are unchanged. Similarly, if initial volatile inventories are restricted to be more Earth-like (<10 Earth oceans, and less $CO_2$ than water by mass), then modern atmosphere distributions barely change (Figure D8).

Rather than attempt to explicitly model nitrogen cycling across diverse planetary conditions, we performed sensitivity tests in which we varied the fixed $N_2$ atmospheric partial pressure over approximately an order magnitude. We find that, for the inner planets, higher $pN_2$ makes atmospheric oxygenation slightly (few percent) less likely, but otherwise changes in background $N_2$ do not affect qualitative results (Appendix D).

## 4. Discussion

### 4.1. Predictions for James Webb Space Telescope Characterization of the Trappist-1 System

The distributions of atmospheric outcomes shown in Figure 6 should not be interpreted literally given the necessarily arbitrary ranges sampled for some underlying parameters, and the possibility of hidden correlations between parameters governing geochemical exchanges or atmospheric escape. However, it is still possible to make qualitative predictions by comparing expected outcomes for different Trappist-1 planets.

In general, we find that abiotic oxygen accumulation becomes less probable with increasing planet–star separation (Figure 10). This trend holds regardless of what threshold is assumed for "oxic," and it is independent of metallic core size, oxygen sink, or initial volatile assumptions. About half of all model runs for 1b and c accumulate potentially detectable amounts of atmospheric oxygen with JWST—around 1 bar $pO_2$ or more (Fauchez et al. 2020b)—after 8 Gyr, whereas 1e, f, and g only possess oxygen-rich modern atmospheres in ~40%, ~20%, and ~10% of model runs, respectively. Trappist-1d is the exception to the general trend, being oxidized in 10%–30% of model runs. The small size of 1d inhibits atmospheric oxygenation: if calculations are repeated assuming 1d is one Earth mass, then it falls along the trend of oxygenation in Figure 9. The only somewhat outlying sensitivity test is the waterworlds case for the outer planets. As discussed above, if surface volatile inventories are large, then oxygen sinks are suppressed, and oxygen that accumulates during the pre-main sequence is more likely to persist.

These findings are broadly consistent with previous models of magma ocean redox evolution. Wordsworth et al. (2018) coupled magma ocean evolution, hydrogen and oxygen photochemistry and radiative transfer, and the XUV evolution of late M dwarfs to make predictions of pre-main sequence oxygen buildup. They found the potential for abiotic buildup was "medium" for Trappist-1b, c, d, and e, and "low" for 1f and g. Similarly, Barth et al. (2021) considered the steam atmosphere, magma ocean evolution of planets 1e, f, and g under extreme internal heating scenarios, and reported that a diversity of final states were possible depending on initial water inventory, including both oxygen-rich and anoxic atmospheres.

With regards to other atmosphere bulk constituents, steam-rich (few % $H_2O$) atmospheres remain a slight possibility for 1b, c, and d, especially in cases where total atmospheric pressure is comparatively low (0.1–10 bar). Atmospheres rich in $CO_2$ and/or $O_2$ remain the most likely outcome, however (Figure 6). For the other planets, detectable quantities of water vapor in the transit chord are unlikely (Figure 11). While Earth-like stratospheric water vapor abundances (few ppm) are unlikely to be detectable with JWST (Fauchez et al. 2019; Komacek et al. 2020), steam-rich upper atmospheres (~0.1%) on the inner planets could be detectable (Lincowski et al. 2018; Lustig-Yaeger et al. 2019a). The bulk atmospheres of 1e, f, and g are also likely to be $CO_2$-dominated or $O_2$-dominated, if a dense secondary atmosphere has been retained.

The likelihood of dense, $CO_2$-rich atmospheres on the inner planets motivates further consideration of Venus analogs as a class of objects (Kane et al. 2014). Moreover, testable predictions for abiotic oxygen accumulation on these planets, whether confirmed or disproved, will inform understanding of oxygen biosignature false-positive mechanisms (Meadows et al. 2018). The strong possibility of $CO_2$- or $O_2$-rich atmospheres in the Trappist system suggested by our models further motivates attempts to understand self-consistent possible photochemical-climate states (Lincowski et al. 2018) and 3D climates (Fauchez et al. 2020a; Turbet et al. 2018; Wolf 2017) of these planets in detail.

Even more broadly, the Trappist-1 system will provide an opportunity to test theoretical predictions on secondary atmosphere loss and retention. An empirical "cosmic shoreline" separates known planetary bodies with and without atmospheres, and can be extrapolated to make predictions for exoplanets given different dominant loss mechanisms (Zahnle & Catling 2017). The PACMAN model in its current form is not an ideal tool for anticipating the presence or absence of a substantial atmosphere given the crude treatment of nonthermal escape and the imposition of a constant $N_2$ background. However, we can make some broad statements on the frequency of atmospheric retention given assumed atmosphere-exchange processes, a range of initial inventories, and constraints on current surface volatile inventories from mass–radius measurements. Specifically, if the combined partial pressure of $CO_2$, $O_2$, and surface water (all phases) after 8 Gyr of evolution is <1 bar, then we presume the atmosphere could be susceptible to complete erosion if nitrogen cycling and escape were explicitly modeled. When large metallic cores are permitted (nominal model), the percentage of model runs with final atmospheres susceptible to complete erosion for 1b, c, and d is 26%, 22%, and 25%, respectively. If solar-system-like metallic core fractions are assumed (<32.5% core mass fraction), then these percentages grow to 56%, 50%, and 30%, respectively. In contrast, <1% of model runs for 1e, f, and g are susceptible to complete atmospheric erosion of all surface volatiles. This outer planet atmospheric loss percentage increases only slightly to 2%–3% if Earth-like initial volatile inventories are assumed (i.e., <10 oceans and less $CO_2$ than water). It should be noted, however, that for the outer planets, liquid water, ice, and dissolved $CO_2$ all count toward surface volatile inventories. If only atmospheric volatiles are considered, then 38%–50% of model runs for 1e, f, and g possess <1 bar $H_2O+O_2+CO_2$ after 8 Gyr. In summary, if the core mass fractions of 1b, c, and d are Earth-like then it would be unsurprising to find that their atmospheres are completely eroded. In contrast, the complete erosion of atmospheres on 1e, f, and g is unlikely, unless virtually all surface volatiles are frozen or if initial volatile inventories were at the low end of our assumed ranges. This provides some reason for optimism that upcoming JWST observations could yield atmospheric detections within the Trappist-1 system.

### 4.2. Caveats and Directions for Future Modeling

The modeling effort described above attempts to represent all important processes that modulate bulk atmosphere evolution. While the model is adequate for reproducing the known evolution of solar-system terrestrials, there are many ways in which planets around late M dwarfs such as Trappist-1 could diverge from their solar-system analogs. If the predictions for Trappist-1 described above differ dramatically from reality, then the following model assumptions ought to be revisited.

First, the model assumes primary atmospheres are short lived and do not impact subsequent evolution. While this is likely a reasonable assumption for solar-system terrestrial planets (Zahnle et al. 2010), many short-period terrestrial exoplanets form with substantial and long-lived $H_2$-dominated atmospheres (Van Eylen et al. 2018). It is not known whether the Trappist-1 planets formed with substantial primary atmospheres. The disks around low-mass stars are typically small, which makes it more challenging to form massive enough cores to trigger gas accretion before nebula dissipation. Formation models for Trappist-1 yield primary atmosphere mass fractions and lifetimes that are comparatively short (Hori & Ogihara 2020). If present, however, such primary atmospheres could impact secondary atmosphere evolution; long-lived $H_2$ atmospheres could result in the escape of heavier volatiles via hydrodynamic drag (Kite & Barnett 2020). The equilibrium temperatures of the Trappist-1 planets are likely too low for this process to inevitably lead to complete desiccation (Kite & Barnett 2020), but the possible impact of primary atmospheres on Trappist-1 planetary redox evolution have yet to be explored with a coupled model.

Second, our modeling approach assumes that carbon, hydrogen, and oxygen are the most important volatile species for planetary redox evolution. While sulfur cycling may have played an important role in the oxygenation of Earth's atmosphere (Gaillard et al. 2011; Olson et al. 2019), sulfur redox fluxes are small compared to the potential for several 100 bar $O_2$ on Trappist-1 planets and will likely only slightly modulate oxygen sinks (Krissansen-Totton et al. 2021b). Similarly, while atmospheric nitrogen is important for providing a noncondensible cold trap for water (Wordsworth & Pierrehumbert 2014), our model results are not sensitive to modest changes in background nitrogen (see Appendix D). However, whether accretion and core formation can yield relative inventories of carbon, hydrogen, oxygen, sulfur, and nitrogen that are radically different to the Earth is an area of active study (Grewal et al. 2019a, 2019b); higher mantle abundances of sulfur and nitrogen could modify planetary redox evolution, whereas very high carbon concentrations could modify mantle geodynamics (Unterborn et al. 2014).

Third, the PACMAN model does not include any explicit photochemistry; it tracks fluxes of oxygen into/out of the combined atmosphere-ocean reservoir, and all outgassed reductants are assumed to instantaneously deplete atmospheric oxygen. This simplification is adequate for estimating long-term oxygen accumulation because if oxygen sources exceed oxygen sinks, then oxidant build-up will occur; neither photolysis reactions nor spontaneous reactions can add net reducing power to the atmosphere-ocean system. (Similarly, neglecting photochemistry likely will not significantly modify bulk atmosphere evolution.) However, our model cannot anticipate photochemical oxygen false positives, such as $O_2$- and CO- rich atmospheres maintained by continuous $CO_2$ photodissociation (Gao et al. 2015; Harman et al. 2015; Hu et al. 2020). While investigations of the role of lightning in such photochemical models (Harman et al. 2018) and improved near-UV water cross sections (Ranjan et al. 2020) may preclude some of these scenarios, photochemical runaways yielding $O_2$–CO-rich atmospheres remain a strong possibility for late M dwarfs such as Trappist-1 (Hu et al. 2020). JWST may be able to distinguish between oxygen produced by secular planetary oxidation and oxygen produced by $CO_2$ dissociation by constraining CO abundances (Krissansen-Totton et al. 2018b; Lustig-Yaeger et al. 2019a). More generally, however, the lack of explicit photochemistry and a generalized radiative transfer scheme precludes exploration of more reducing CO–$H_2O$–$H_2$ atmospheres.

It should also be noted that there are simplifications in the physics represented in the PACMAN model that could lead to errant predictions. The model does not accommodate ice albedo feedbacks or $CO_2$ ice condensation, the latter of which has been shown in 3D circulation models to be potentially important in sculpting the atmospheres of the outer Trappist-1 planets (Turbet et al. 2020b). Atmospheric collapse via $CO_2$ condensation would presumably make hydrogen escape even smaller, thereby making abiotic oxygen accumulation even less likely on the outer planets. We are also assuming orbital evolution has not impacted the atmospheric evolution of the Trappist-1 planets, which is plausible given that the observed orbital resonances are easily disrupted (Raymond et al. 2022).

Strong tidal heating could extend the longevity of terrestrial planet magma oceans if the tidal contribution to the surface energy budget is sizeable (Barnes et al. 2013; Barth et al. 2021). While the latest constraints on the eccentricity of 1b and c are consistent with zero, and $e < 0.01$ for the other Trappist planets (Agol et al. 2021), it remains plausible that tidal heating is the largest source of internal heat for the inner planets (Dobos et al. 2019; Turbet et al. 2018). With that said, even under most extreme heating tidal heating scenario (i.e., $\sim 10^4$ TW = $\sim 20$ W m$^{-2}$), the tidal heating contribution (Dobos et al. 2019) to the surface energy budget is small compared to our assumed uncertainty in albedo, a factor of 2 in absorbed stellar radiation. Whether such excessive tidal heating affects interior evolution depends on the locus of heating, that is, whether the dissipation of tidal heat drives mantle convection and melt production (and therefore enhance oxygen sinks), or merely heats the surface volatile reservoir. We have crudely accounted for modest tidal heating by adopting a high estimate for the core heat flow and two orders of magnitude span in radionuclide inventories, but one opportunity for future work would be to fully couple tidal-atmosphere–interior interactions to test the impact of different orbital evolution scenarios on atmospheric composition. Notably, the timescale over which tidal heating evolves is uncertain (Becker et al. 2020), and could differ from that of radionuclide decay.

Finally, the treatment of atmospheric escape in the PACMAN model necessarily involves many simplifications. We are not self-consistently solving the hydrodynamic equations given atmospheric composition and incoming spectral energy distribution, but are instead adopting approximate analytic expressions for XUV-driven and diffusion-limited loss (see Appendix A) and sampling a broad range of governing parameters (e.g., escape efficiency, thermosphere and cold-trap temperatures, etc.) This parameterization ought to encompass a very broad range of escape scenarios, but it could overlook key physical dependencies and feedbacks which would restrict our probability distribution of outcomes. We have also assumed nonthermal escape is nonfractionating, which is appropriate for impact erosion but not for stellar wind erosion or other photochemical loss mechanisms. Coupling atmosphere–interior models such as PACMAN to physically motivated escape models is a promising area for future work.

## 5. Conclusions

Whether the Trappist-1 planets have accumulated and retained oxygen-rich atmospheres due to extensive hydrogen loss is challenging to anticipate. For the inner planets (1b and c), oxygen accumulation depends primarily on the initial water endowment and the efficiency of dry crustal sinks. Detectable oxygen ($>1$ bar) persists in the modern atmosphere approximately half the time for the inner planets 1b and c. In contrast, abiotic oxygen accumulation is less probable for the outer planets (1e, f, and g). For the outer planets, atmospheric oxygenation depends not only on initial water inventory but also on atmospheric escape physics, initial mantle redox state, and the efficiency of several other crustal oxygen sinks such as water–rock reactions and extrusive lava oxidation. While the probability of oxygen-rich modern atmospheres decreases with orbital separation, this multiplicity of important influences means observed patterns of atmospheric oxygenation may evade simple explanations.

Predicting the presence or absence of substantial atmospheres is fraught, especially without more nuanced models of atmospheric nitrogen and nonthermal escape processes. Broadly speaking, however, the lack of any substantial atmosphere would be unsurprising on the inner planets. This is because, of the PACMAN model runs that are compatible with observed mass–radius constraints, only about half retain a dense atmosphere ($pCO_2 + pH_2O + pO_2 > 1$ bar) after 8 Gyr. Complete atmospheric erosion is far less likely on the outer planets given that the lifetimes of (easily eroded) runaway greenhouse atmospheres are limited, and because mass–radius constraints do not restrict condensed surface volatile reservoirs, and so more model runs with substantial atmospheres are compatible with observed densities.

If the Trappist planets did retain substantial atmospheres, then they are likely to be either $CO_2$-domainted or $CO_2$–$O_2$ dominated (or CO–$O_2$ dominated if there is substantial photochemical dissociation of $CO_2$). While atmospheres rich in water vapor cannot be completely ruled out for the inner planets, the low instellations received by the outer planets suggests that atmospheric water vapor is either absent or will not be detectable in transit spectra. The amount of atmospheric $CO_2$ in modern atmospheres depends primarily on initial endowments, as well as escape physics (for the inner planets) and carbon cycle feedbacks (for the outer planets).

Finally, there are many simplifications in the PACMAN model that could produce incorrect predictions. The neglect of primary atmospheres, the absence of explicit photochemistry, simplified escape parameterizations, and the restriction of volatile cycling to C-, H-, and O-bearing species could all be potentially problematic. However, any mismatches between the predictions above and upcoming JWST observations will motivate future model development and improved theoretical understanding.

We thank the two anonymous reviewers for their helpful and constructive comments. Joshua Krissansen-Totton was supported by the NASA Sagan Fellowship and through the NASA Hubble Fellowship grant No. HF2-51437 awarded by the Space Telescope Science Institute, which is operated by the Association of Universities for Research in Astronomy, Inc., for NASA, under contract NAS5-26555. Jonathan Fortney was supported by an Investigator grant from the Simons Foundation and NASA's Interdisciplinary Consortia for Astrobiology Research (NNH19ZDA001N-ICAR) under award number 19-ICAR19_2-0041.

## Appendix A
## PACMAN Model Description

### A.1. Atmospheric Escape and Trappist-1 Stellar Evolution

Figure A1 shows the assumed time evolution of Trappist-1 bolometric luminosity (Baraffe et al. 2015) and XUV luminosity, with uncertainty (Birky et al. 2021). Atmospheric thermal escape is assumed to be either diffusion limited or XUV limited, depending on atmospheric composition and the incident stellar XUV flux. In the diffusion limit, we assume that eddy diffusion dominates vertical transport at altitudes where water is more abundant than atomic H and O (Catling & Kasting 2017), and so escape of hydrogen is limited by the diffusion of atomic hydrogen through the background atmosphere:

$$\begin{aligned}
\varphi_{\text{diff}} &= b_H f_H (1/H_n - 1/H_H) \\
b_H &= \frac{b_{H-CO_2} pCO_2 + b_{H-N_2} pN_2 + b_{H-O} pO}{pCO_2 + pN_2 + pO} \\
H_H &= \frac{8.314 T_{\text{cold-trap}}}{\mu_H g} \\
H_n &= \frac{8.314 T_{\text{cold-trap}}}{\bar{\mu} g}.
\end{aligned} \quad (1)$$

Here, $b_{i-j}$ (mol/m/s) is the binary diffusion coefficient of the $i$th species through the $j$th species (Marrero & Mason 1972; Zahnle & Kasting 1986). These are weighted by the thermosphere mixing ratios of each noncondensible constituent ($CO_2$, $N_2$, and atomic O), which are obtained from the upper-atmosphere mixing ratios, and from assuming that all water is dissociated. The scale height of hydrogen, $H_H$ (m), and of the background gases, $H_n$ (m), depend on upper-atmosphere temperature, $T_{\text{cold-trap}}$ (see main text). The hydrogen mixing ratio in the thermosphere, $f_H$, is assumed to be double the upper-atmosphere water mixing ratio, $f_{H_2O}$. The diffusion-limited escape flux is $\varphi_{\text{diff}}$ (mol H/m²/s).

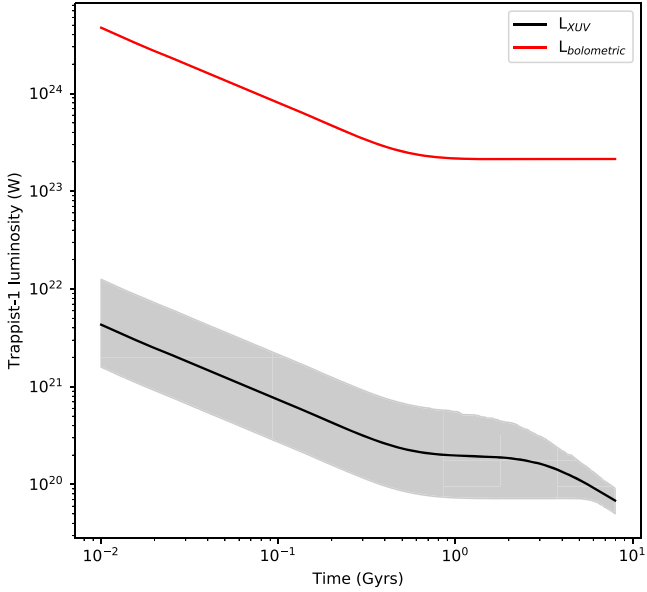

**Figure A1.** Assumed evolution of Trappist-1 bolometric luminosity (red line) and XUV luminosity (black line with 95% confidence shaded gray).

To calculate the XUV-driven hydrodynamic escape of hydrogen, and associated oxygen and $CO_2$ drag, we follow Odert et al. (2018) and Zahnle & Kasting (1986). The total XUV energy mass-loss rate, $\Phi_{XUV}$ (kg/m²/s) is specified by the following equation:

$$\Phi_{XUV} = \frac{\varepsilon(F_{XUV}, X_O, \zeta, \varepsilon_{lowXUV}) F_{XUV} r_p}{4GM_p}. \quad (2)$$

Here, $F_{XUV}$ is the XUV flux (W/m²) received from Trappist-1 (Figure A1). The efficiency of hydrodynamic escape, $\varepsilon$, is a function of atmospheric composition and XUV stellar flux (Krissansen-Totton et al. 2021b). In general, the XUV-driven mass flux will be partitioned between hydrogen loss, oxygen drag and, under very high XUV fluxes, $CO_2$ drag (Krissansen-Totton et al. 2021b; Odert et al. 2018). For example, the oxygen fractionation factor, $\chi_O$, can be obtained:

$$\chi_O = 1 - \frac{g(m_O - m_H)b_{H-O}}{\Phi_H k_B T_{thermo}(1 + X_O)}. \quad (3)$$

Here, $T_{thermo}$ is the thermosphere temperature described in the main text, $k_B$ is the Boltzmann constant, $m_i$ (kg) is the mass of the $i$th species, and $X_O$ is the upper-atmosphere mixing ratio of oxygen. The hydrogen escape flux, $\Phi_H$ (molecules H/m²/s), can be obtained by analytically solving Equations (4), (5), and (6) in Odert et al. (2018).

The combination of diffusion-limited and XUV-limited thermal escape parameterizations ensures diffusion-limited hydrogen escape for low stratospheric abundances, and a smooth transition to XUV-driven escape as the upper atmosphere becomes steam dominated (Krissansen-Totton et al. 2021b). The precise transition abundance is unknown and will, in general, depend on conductive and radiative cooling of the upper atmosphere as well as downward diffusive transport. Here, it is represented by the free parameter, $\lambda_{tra}$, which ranges from $10^{-2}$ to $10^2$ and is sampled uniformly in log space. The efficiency of hydrodynamic escape is parameterized by loosely following the approach of Wordsworth et al. (2018). If the XUV stellar flux is insufficient to drag oxygen, then the efficiency is equal to a constant, $\varepsilon_{lowXUV}$, which is randomly sampled from 1% to 30%. Alternatively, if the XUV stellar flux exceeds what is required to drag oxygen, then some portion of the excess energy, $\zeta_{high}$, goes into driving further escape, whereas the rest, $1 - \zeta_{high}$, is assumed to be efficiently radiated away. The efficiency factor, $\zeta_{high}$, is randomly sampled from 0%–100% for complete generality. See Krissansen-Totton et al. (2021b) for analytic expressions.

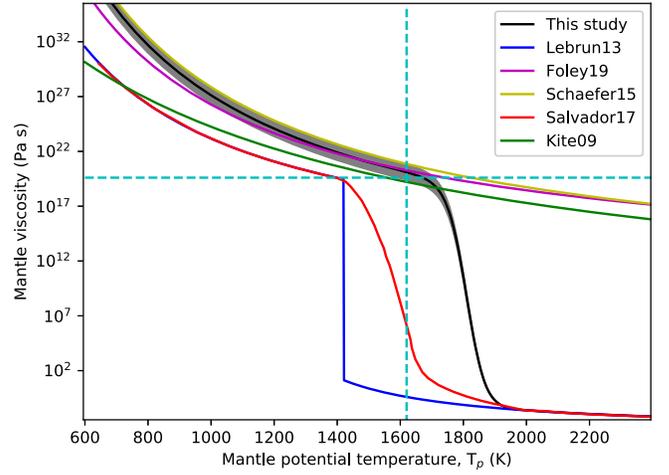

**Figure A2.** Assumed viscosity parameterizations compared to other parameterizations from the literature (Foley 2019; Kite et al. 2009; Lebrun et al. 2013; Salvador et al. 2017; Schaefer et al. 2016). The dashed cyan lines represent the mantle potential temperature and viscosity required to reproduce the modern Earth's melt production and plate velocity. Adopted from Krissansen-Totton et al. (2021b).

Rather than attempt to explicitly model the myriad nonthermal escape processes and their uncertain functional and time dependencies, we instead assume a constant rate of loss totaling 1–100 bar over the 8 Gyr lifetime of Trappist-1 (sampled log-uniformly). This range is loosely based upon models of ion escape rates for planets around Trappist-1 (Dong et al. 2018) and Proxima Centauri (Garcia-Sage et al. 2017). For simplicity, atmospheric constituents are not fractionated by nonthermal escape, meaning loss rates of $CO_2$, $O_2$, and $H_2O$ are weighted by mixing ratio. Our chosen range for nonthermal escape does not encompass extremely high impact erosion rates that have been proposed (Kral et al. 2018), but dynamical stability considerations disfavor such high rates of impact erosion post-nebula dissipation (Raymond et al. 2022).

### A.2. Interior Evolution

The interior evolution model is identical to that described in Krissansen-Totton et al. (2021b) except for the following changes. As noted in the main text, the initial radiogenic inventory is sampled from a range that spans two orders of magnitude to accommodate potentially large contributions from tidal heating. Additionally, mantle volatile retention during magma ocean solidification is included by accounting for melt compaction in the freezing front. Following Hier-Majumder & Hirschmann (2017) we used the following expression to determine the melt fraction trapped in the mantle

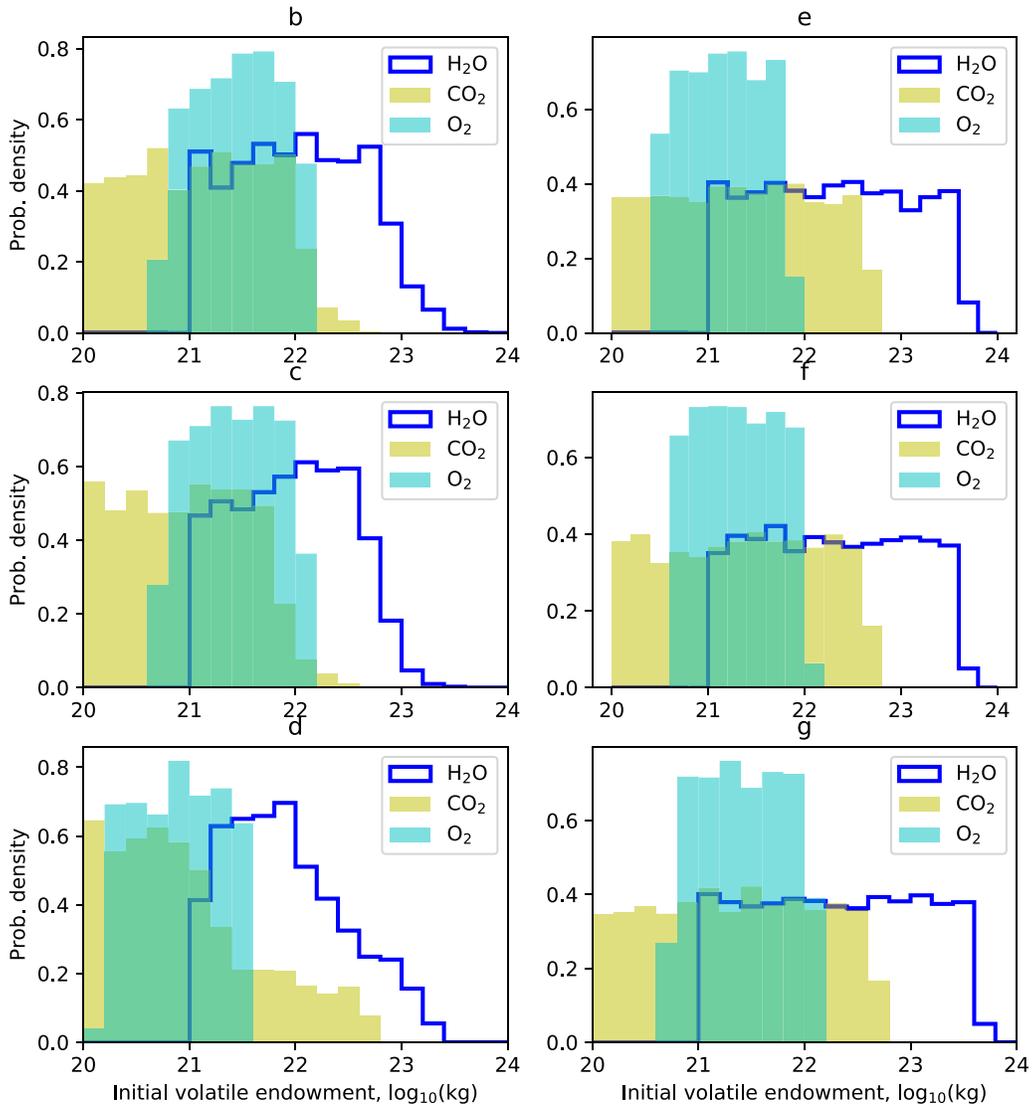

**Figure A3.** Initial volatile inventories for planets 1b–g for all model runs that satisfy modern mass–radius constraints. Yellow shaded distributions denote initial $CO_2$, cyan shaded distributions denote initial free oxygen, and blue lined distributions denote initial water inventories. The strong constraints on modern surface volatile inventories for the inner planets imply they were unlikely to have formed with more than ∼100 Earth oceans, whereas initial water inventories for the outer planets could have exceeded 100 Earth oceans.

as the solidification front moves toward the surface, $f_{TL}$:

$$f_{TL} = -\frac{0.3\tau_c}{T_{Liquidus} - T_{Solidus}}\frac{dT_P}{dt}. \quad (4)$$

Here, $\tau_c = 10^6$ years is the assumed compaction timescale, which is derived from a ∼cm/yr characteristic velocity of matrix sedimentation and ∼10 km freezing front (Hier-Majumder & Hirschmann 2017). Rapid magma ocean solidification (i.e., a large rate of change in mantle potential temperature), results in greater volatile retention. Additionally, we require that $f_{TL}$ is bounded below by 0 (negative trapped melt fractions are unphysical) and above by 0.3 (retained melt fraction does not exceed disaggregation melt fraction). The trapped melt fraction is incorporated into the system of equations governing water and carbon dioxide reservoir evolution.

Mantle viscosity is assumed to depend on temperature, as shown in Figure A2 (Krissansen-Totton et al. 2021b), which encompasses both magma ocean and solid-mantle evolution.

### A.3. Initial Volatile Inventories

The initial volatile inventories assumed in nominal calculations are described in Table 1 in the main text. However, only those model runs that satisfy mass–radius constraints were plotted in Figures 2–5 and used in the subsequent analyses. The corresponding initial volatile inventories for these "successful" model runs are shown in Figure A3.

### Appendix B
### Atmosphere–Interior Calculations for Trappist-1c, d, and f

Figure B1 shows the atmosphere–interior evolution of Trappist-1c, as calculated using the PACMAN model. Only model runs that satisfy mass–radius constraints (volatile mass fraction <0.08%) are shown. The outputs are qualitatively similar to 1b.

Figure B2 shows the atmospheric evolution of Trappist-1f, as calculated using the PACMAN model. Only model runs that satisfy mass–radius constraints (volatile mass fraction <14.0%) are shown. The model outputs are qualitatively similar to 1g, except that surface temperatures are typically warmer.

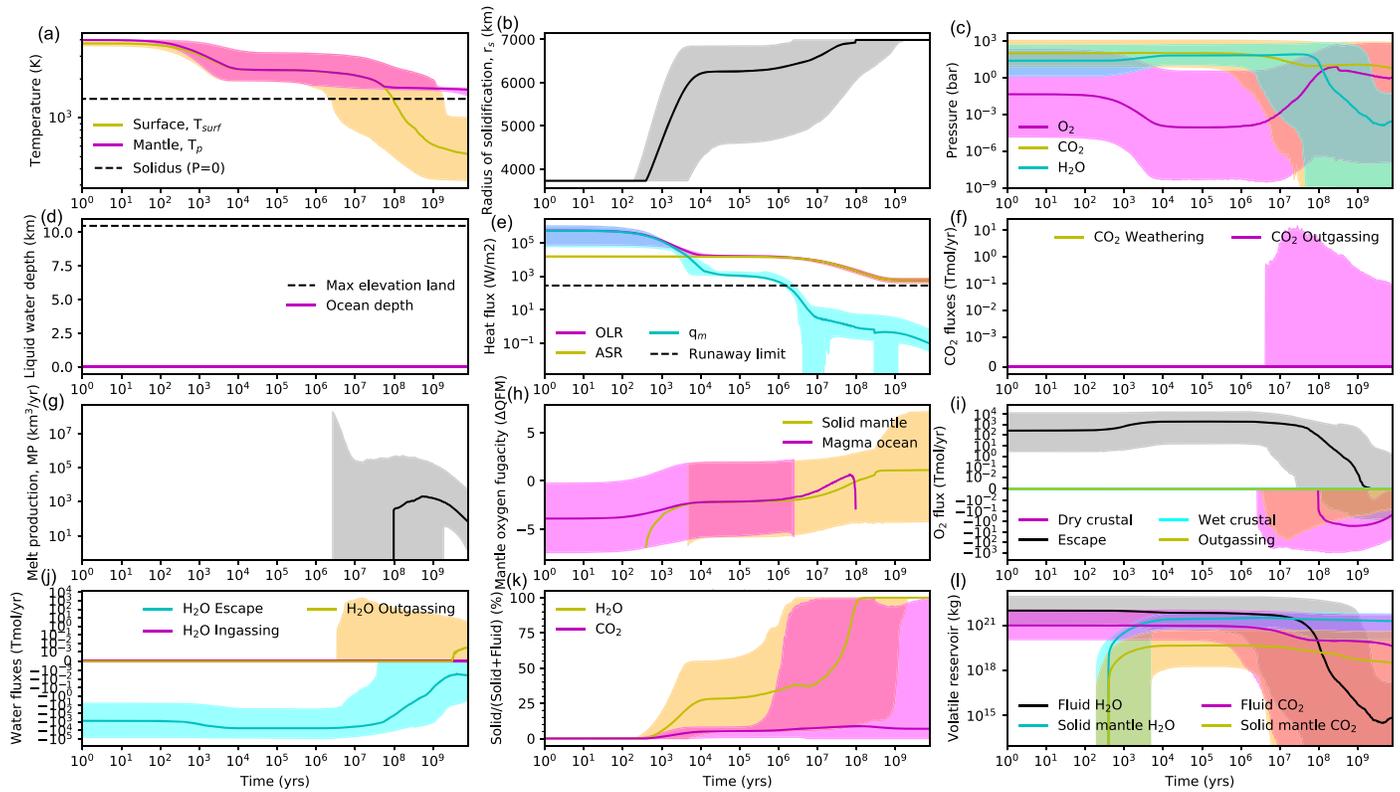

**Figure B1.** Trappist-1c predicted atmosphere–interior time evolution. The lines are median values and shaded regions denote 95% confidence intervals. The time evolution spans from post-accretionary magma ocean to the present day ($8 \times 10^9$ yr). Subplot (a) shows the evolution of mantle potential temperature (magenta) and surface temperature (orange) alongside the silicate solidus (black dashed line). The magma ocean (b) persists from anywhere between a few million years to a few billion years. Water catastrophically degasses from the magma ocean and is subsequently lost to space via hydrodynamic escape (c, j). Trappist-1c's small stellar separation means the absorbed shortwave radiation never drops below the runaway greenhouse limit (e), and so liquid water never condenses on the surface (d). Atmospheric oxygen may be produced via escape (c, i), but this oxygen is drawn down by geological sinks (i), namely magmatic outgassing of reduced species (orange) and dry crustal oxidation (magenta). The silicate mantle is gradually oxidized by this net hydrogen loss (h). Volatile cycling is controlled by the rate at which fresh crust is produced (g), and while melt production typically continues throughout Trappist-1c's evolution, outgassing fluxes are usually small because the mantle is left comparatively desiccated after the magma ocean (l).

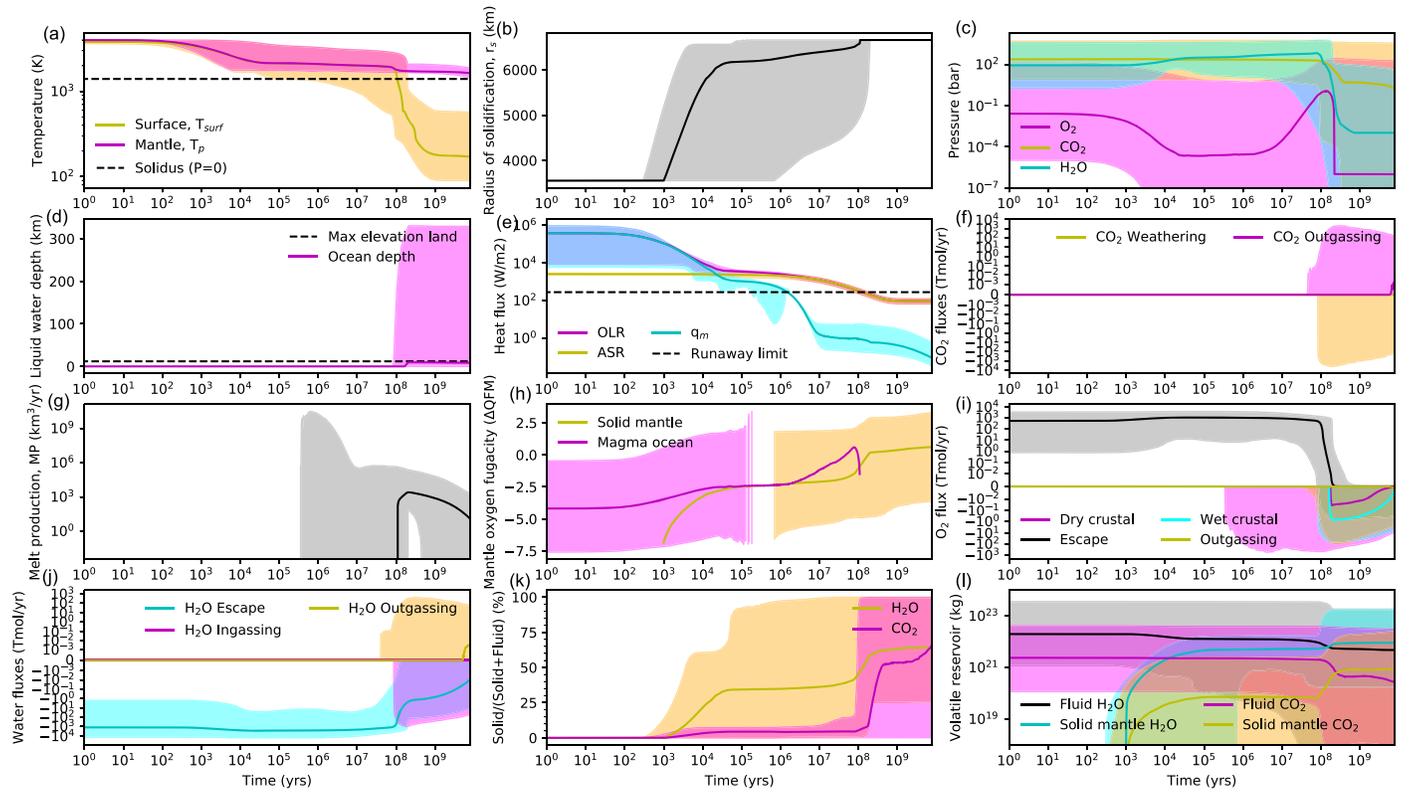

**Figure B2.** Trappist-1f predicted atmosphere–interior time evolution. The lines are median values and shaded regions denote 95% confidence intervals. The time evolution spans from post-accretionary magma ocean to the present day ($8 \times 10^9$ yr). Subplot (a) shows the evolution of mantle potential temperature (magenta) and surface temperature (orange) alongside the silicate solidus (black dashed line). The magma ocean (b) persists from anywhere between a few tens of thousands of years to 100 million years, depending primarily on initial water endowment and albedo. Water catastrophically degasses from the magma ocean and is subsequently lost to space via hydrodynamic escape (c, j). The runaway greenhouse phase ends when Trappist-1 dims sufficiently for absorbed shortwave radiation to drop below the runaway greenhouse limit (e). When this occurs, any remaining atmospheric water vapor (c) condenses onto the surface producing an ocean or frozen crust (d). A temperate carbon cycle also commences (f, k), where volatile cycling is controlled by the rate at which fresh crust is produced (g). Once the steam atmosphere has condensed, atmospheric oxygen produced via escape during the pre-main sequence (c, i) may be drawn down by geological sinks (i): magmatic outgassing of reduced species (orange), water–rock reactions (cyan), and dry crustal oxidation (magenta). In most cases, these sinks are sufficient to remove virtually all oxygen from the atmosphere (c).

## Appendix C
## Variables Controlling Atmospheric $CO_2$ and Water Vapor

Figure C1 shows a correlation plot for atmospheric $CO_2$. Specifically, for each of the Trappist-1 planets b–g, the linear correlation $R^2$ value and $p$-values are plotted for the 26 variables that control atmospheric evolution listed in Table 1. Unsurprisingly, final atmospheric $CO_2$ is strongly correlated with initial $CO_2$ endowment for all planets. Additionally, for the inner planets, final atmospheric $CO_2$ is strongly dependent on the physics of atmospheric escape and stellar evolution. Larger escape efficiencies for XUV-driven thermal escape produce smaller final $CO_2$ inventories, and larger $\zeta_{high}$ values—meaning a greater portion of stellar XUV radiation goes into driving hydrodynamic escape—also yield less atmospheric $CO_2$. Taken together, these three relationships imply higher thermal escape fluxes lead to less $CO_2$, ostensibly due to some combination of $CO_2$ drag in the hydrodynamic outflow (Odert et al. 2018) and thermal escape of hydrogen and oxygen allowing for greater nonthermal escape of $CO_2$ at later times. For 1b, c, and d, atmospheric $CO_2$ is strongly negatively correlated with nonthermal escape.

For the outer planets, atmospheric escape is less important for $CO_2$ abundances. Instead, there is a strong negative correlation with initial water endowment because larger oceans mean more carbon can be dissolved in surface (or subsurface) oceans, thereby reducing the atmosphere inventory. This is also why there is an inverse correlation with alkalinity for the outer planets since higher conservative cation abundances mean more carbon partitions into the oceans. For 1f and g there is also a negative correlation with albedo because higher albedos tend to produce extremely cold climates whereby seafloor weathering eventually removes virtually all carbon from the atmosphere (Nakayama et al. 2019). The opposite relationship holds for 1e: cold climates imply high $CO_2$ is required to balance the carbon cycle (atmospheric $CO_2$ is similarly negatively correlated with the weathering limit for 1e).

Figure C2 shows a correlation plot for atmospheric $H_2O$. For each of the Trappist-1 planets b–g, the linear correlation $R^2$ value and $p$-values are plotted for the 26 variables that control atmospheric evolution. Atmospheric water vapor is positively correlated with initial $CO_2$ for 1d, e, and f. This is because atmospheric $CO_2$ warms the surface climate, therefore resulting in greater atmospheric water vapor (for the inner planets, greater $CO_2$ abundances also throttle water loss). Atmospheric water vapor is similarly negatively correlated with albedo for the outer planets where liquid surface water is permitted because the cold atmospheres produced by high albedos hold less water vapor (1g is essentially frozen regardless of albedo).

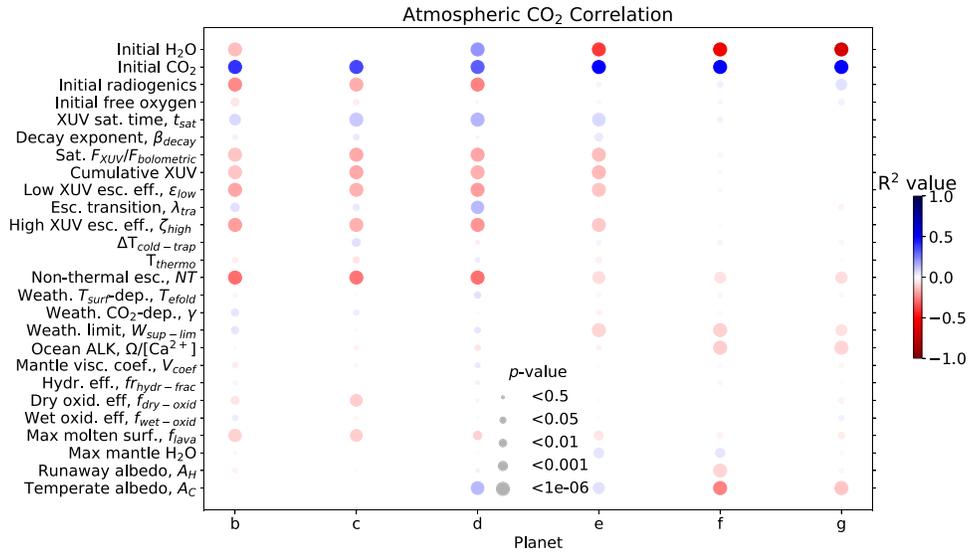

**Figure C1.** Linear correlations between atmospheric $CO_2$ after 8 Gyr of evolution and the input parameters from Table 1 (rows). Columns denote results for Trappist-1 planets 1b, c, d, e, f, and g. The color of each dot shows the $R^2$ value of the linear correlation, whereas the size of the dot reflects the statistical significance of the relationship (larger dots imply a more significant linear relationship). For the outer planets, atmospheric $CO_2$ depends primarily on initial volatile inventories, dissolution of carbon in surface oceans, and albedo (which modulates climate and therefore carbon cycling). For the inner planets, atmospheric $CO_2$ is additionally sculpted by thermal and nonthermal escape processes and stellar evolution.

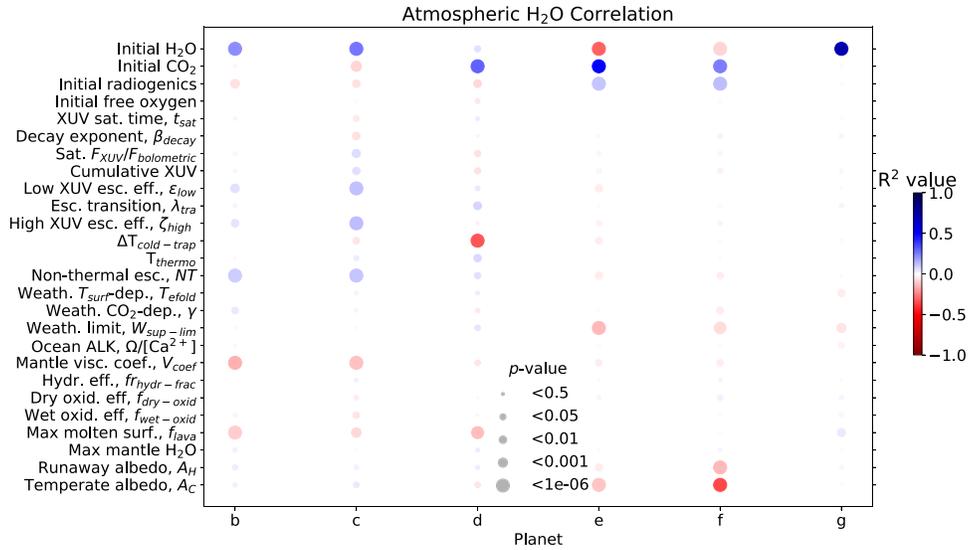

**Figure C2.** Linear correlations between atmospheric water vapor after 8 Gyr of evolution and the input parameters from Table 1 (rows). Columns denote results for Trappist-1 planets b, c, d, e, f, and g. The color of each dot shows the $R^2$ value of the linear correlation, whereas the size of the dot reflects the statistical significance of the relationship (larger dots imply a more significant linear relationship).

Somewhat surprisingly, there is no obvious pattern of positive correlation with initial water inventory. There is even a negative correlation for 1e and f. This occurs because greater water inventories imply more dissolution of carbon in the oceans, and thus cooler climates and less atmospheric water vapor. The positive correlation for 1g is ostensibly because, for such cold climates, the only way to retain atmospheric $CO_2$ and water vapor is to have such a large water inventory that seafloor weathering stops due to the pressure overburden suppression of crustal production.

## Appendix D
## Sensitivity Tests

### D.1. Small Metallic Core Sensitivity Test

Figure D1 is identical to Figure 6 in the main text except that it shows probability distributions from the sensitivity test for a smaller range of metallic core mass fractions (0%–32.5%), as opposed to the 0%–50% range in the nominal model. Figure D2 shows the correlation plot for the same sensitivity test (otherwise identical to Figure 8 in the main text). Broadly speaking, restricting the range of permitted core mass fractions does not change qualitative results. Figure D3 is identical to

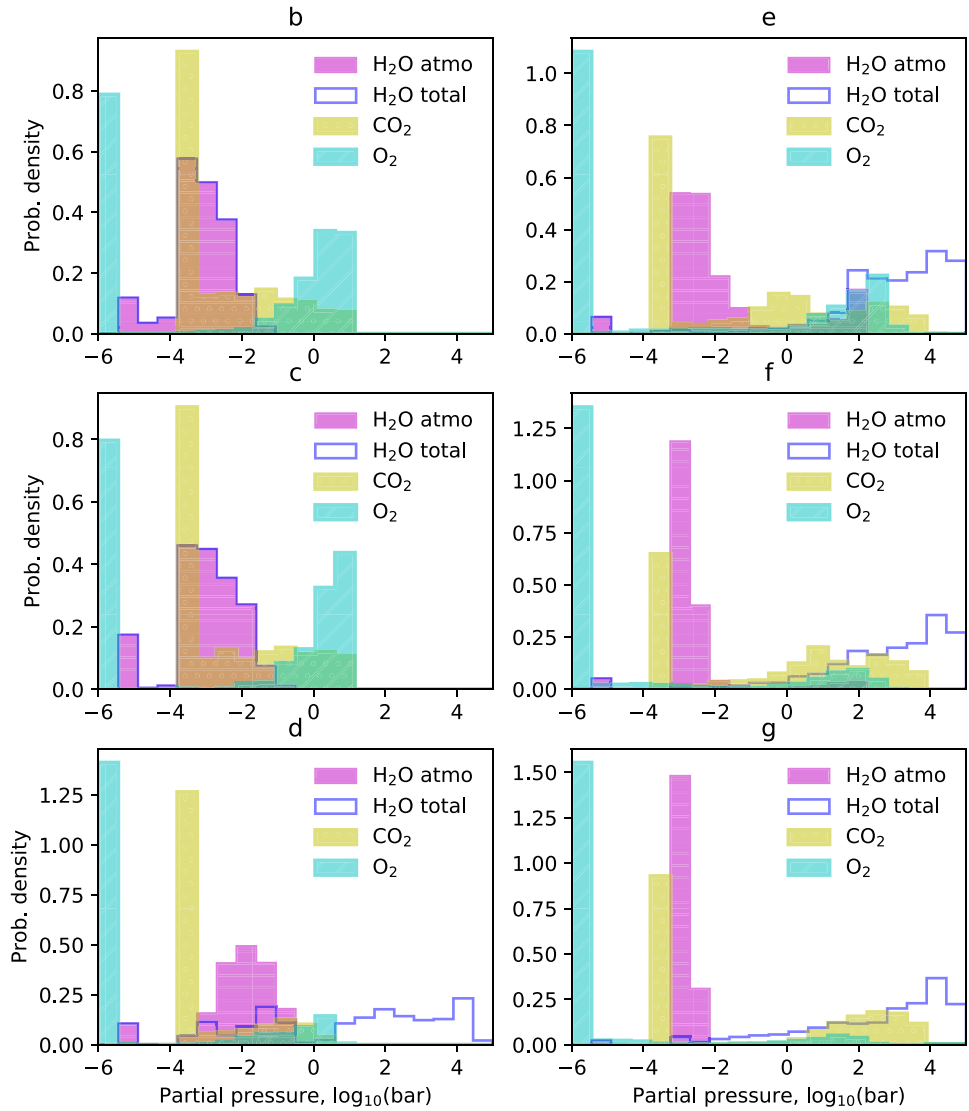

**Figure D1.** Probability distributions for modern atmospheric abundances of Trappist-1 planets for the smaller core mass fraction sensitivity test, as predicted by the PACMAN geochemical evolution model. Partial pressure distributions for carbon dioxide (yellow), oxygen (cyan), and water vapor (magenta) are shown for 1b, c, d, e, f, and g. The total surface water inventory is denoted by the unfilled, blue line distribution; this is identical to the water vapor distribution for the inner planets, but not for the outer planets because surface water can condense as a liquid or solid. Note that there are numerical cutoffs for carbon dioxide, oxygen, and water vapor at $10^{-4}$, $10^{-6}$, and $10^{-6}$, respectively.

Figure A3 except that it shows initial volatile inventories from the small core mass fraction sensitivity test. Here, we find that a smaller core mass fraction implies less initial volatile inventories for the inner planets, but does not affect the outer planets appreciably.

### D.2. Outer Planets as Waterworlds

Figures D4, D5, and D6 show the waterworld sensitivity tests for Trappist-1e, f, and g, respectively.

### D.3. Applying Venus Constraints on Crustal Oxidation Efficiency

Figure D7 is identical to Figure 6 in the main text except that it shows results from the sensitivity test where dry crustal oxidation is efficient, similar to that of Venus, $f_{\rm dry-oxid} > 0.1\%$ (Krissansen-Totton et al. 2021a). A 0%–32.5% core mass fraction range is also assumed. Results are qualitatively similar to Figure D1, except that oxygen-rich atmospheres are slightly less probable.

### D.4. Earth-like Initial Volatile Inventories

Figure D8 is identical to Figure 6 in the main text except that it shows results from the sensitivity test where initial volatile inventories are restricted to Earth-like ranges, 1–10 Earth oceans of water and less $CO_2$ than water (by mass). A 0 to 32.5% core mass fraction range is also assumed. The qualitative trend likelihood of abiotic oxygenation is unchanged.

### D.5. Different Background Nitrogen Inventories

Figure D9 is analogous to Figure 10 in the main text, except that we are showing the impact of the background $N_2$ pressure on the likelihood of abiotic oxygenation. For the inner planets 1b and c, higher $pN_2$ results in slightly less frequent atmospheric oxygenation. This is attributable to smaller

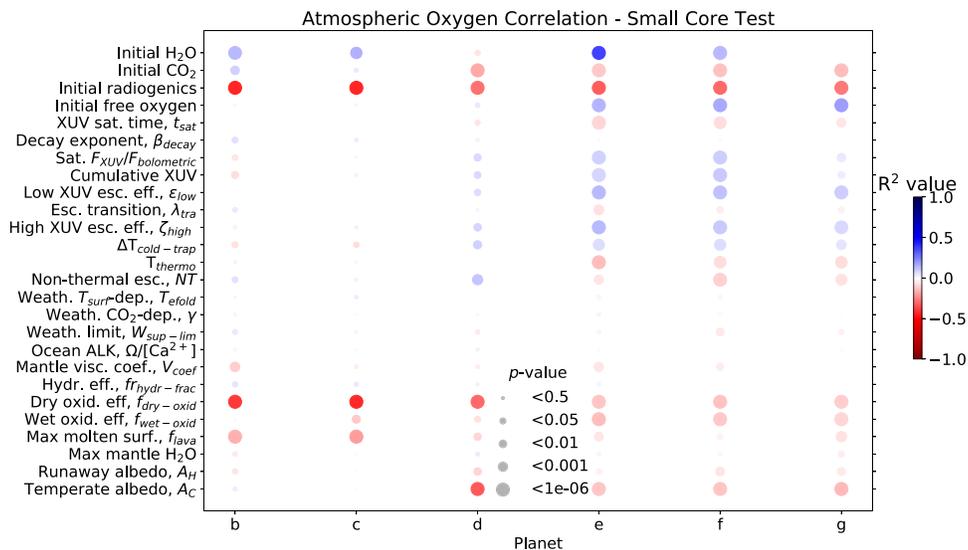

**Figure D2.** Linear correlations between atmospheric oxygen after 8 Gyr of evolution and the input parameters from Table 1 for smaller core mass fraction sensitivity test. Columns denote results for Trappist-1 planets b, c, d, e, f, and g. The color of each dot shows the $R^2$ value of the linear correlation, whereas the size of the dot reflects the statistical significance of the relationship (larger dots imply a more significant linear relationship).

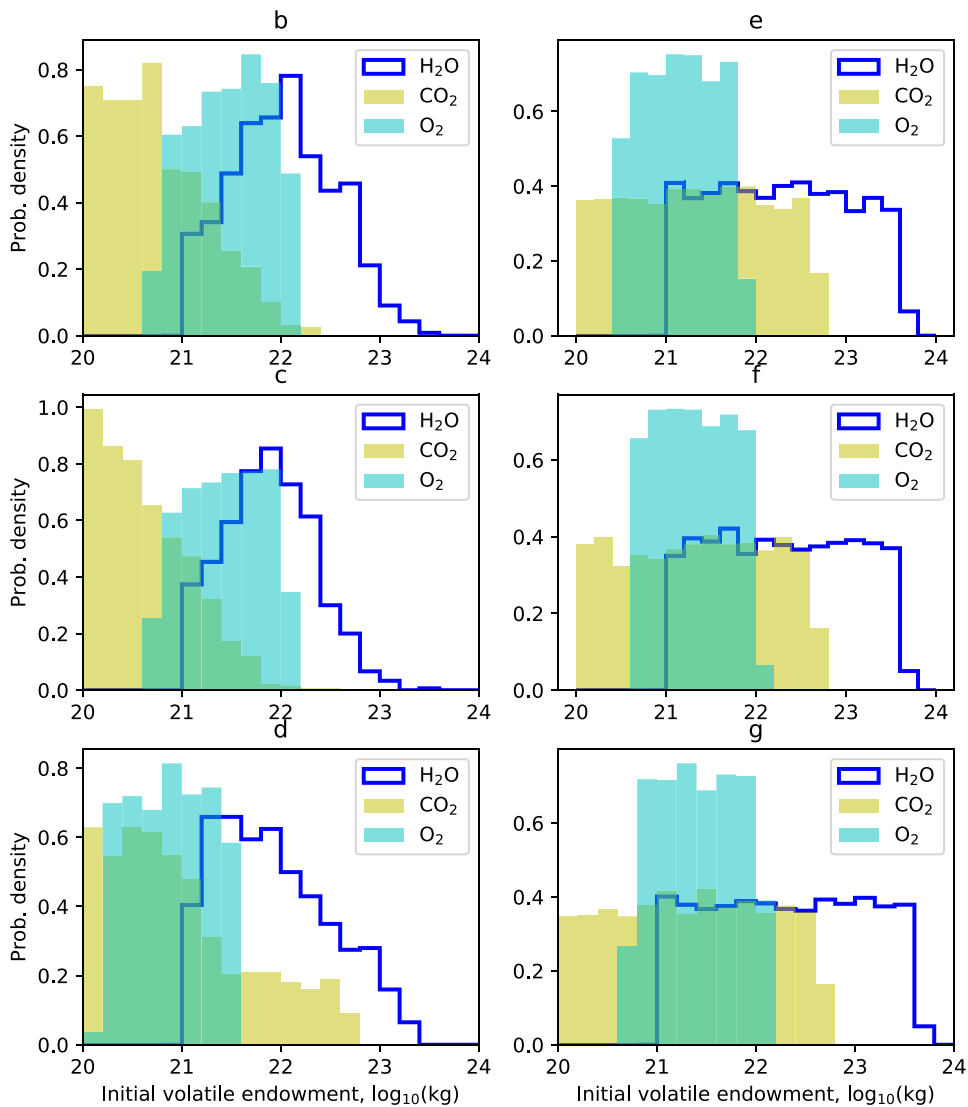

**Figure D3.** Initial volatile inventories for planets 1b–g for all model runs that satisfy modern mass–radius constraints for low core mass fraction sensitivity test. Yellow shaded distributions denote initial $CO_2$, cyan shaded distributions denote initial free oxygen, and blue lined distributions denote initial water inventories.

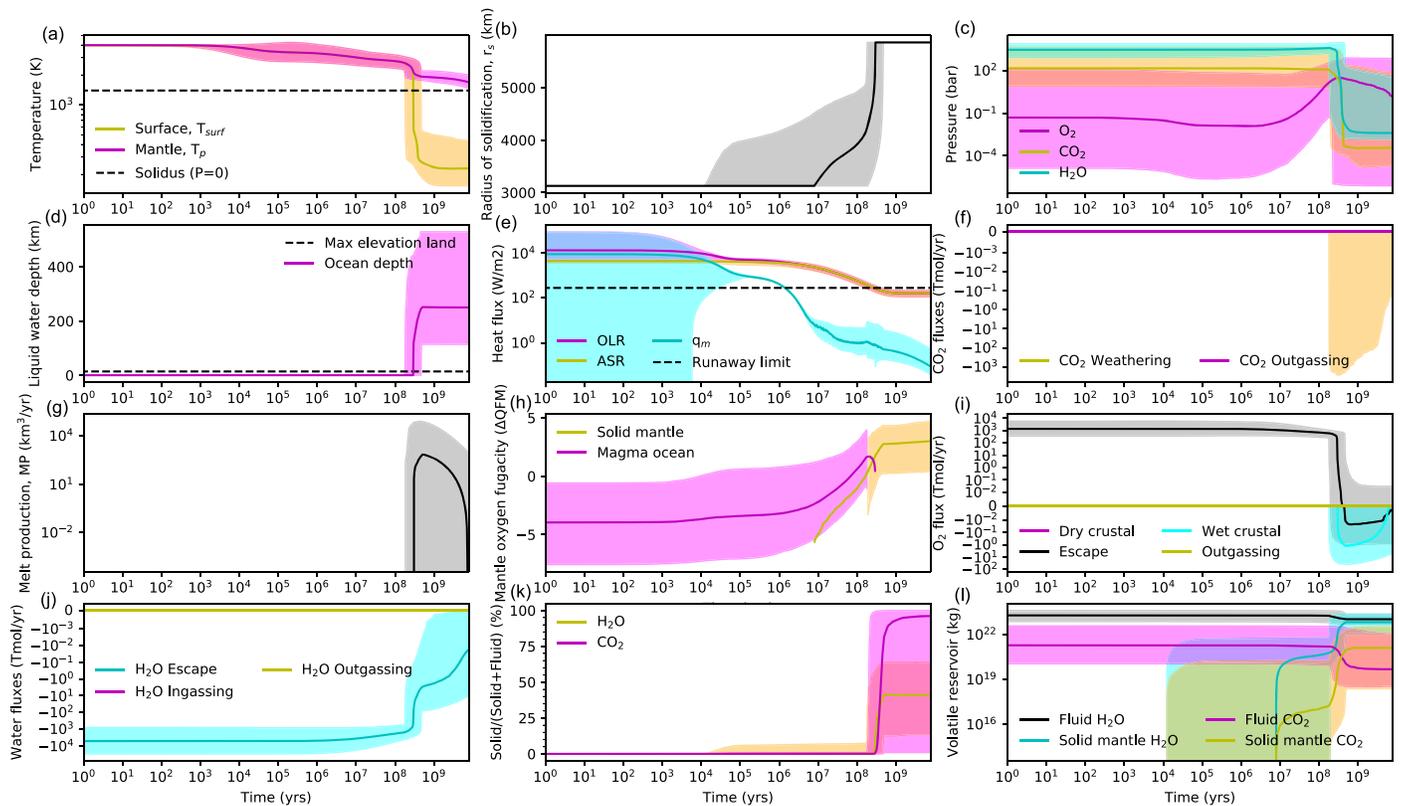

**Figure D4.** Trappist-1e predicted atmosphere–interior time evolution from waterworld sensitivity test. In this calculation, only model runs that terminate with a water mass fraction >1.1% (and <11.6%) are plotted, representing a scenario where the planet's core mas fraction exceeds 25% (and is <50%), meaning a large surface volatile inventory is required to account for observed density. The lines are median values and shaded regions denote 95% confidence intervals. The time evolution spans from post-accretionary magma ocean to the present day ($8 \times 10^9$ yr). Subplot (a) shows the evolution of mantle potential temperature (magenta) and surface temperature (orange) alongside the silicate solidus (black dashed line). The magma ocean (b) persists for hundreds of millions of years due to the large assumed initial water inventory. Water that degasses from the magma ocean is subsequently lost to space via hydrodynamic escape (c, j). The runaway greenhouse phase ends when Trappist-1 dims sufficiently for absorbed shortwave radiation to drop below the runaway greenhouse limit (e). When this occurs, all remaining atmospheric water vapor (c) condenses onto the surface producing an ocean (d). Subsequently, the overburden pressure from surface volatiles suppresses crustal oxygen sinks (f, i). Atmospheric oxygen produced via escape during the pre-main sequence (i, j) is more likely to persist over 8 Gyr (c).

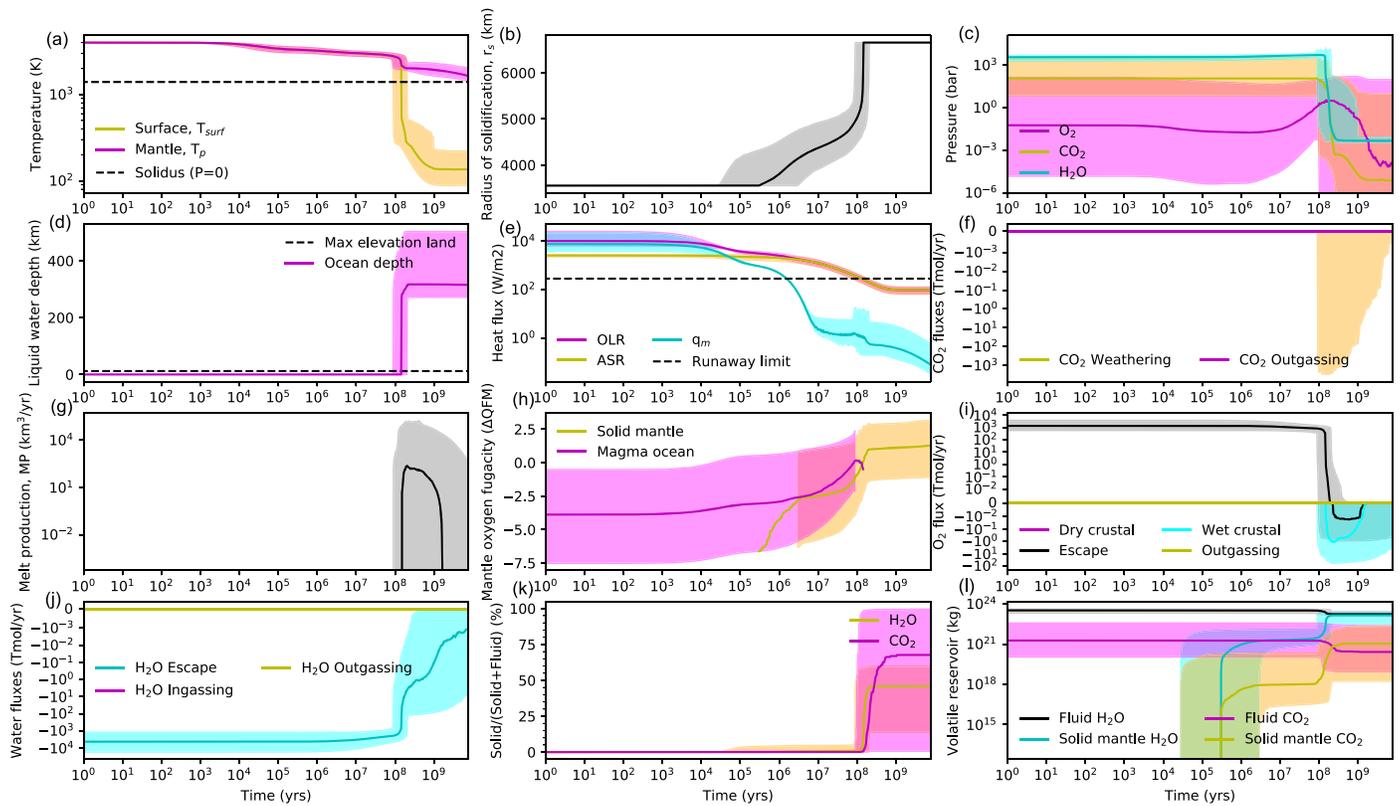

**Figure D5.** Trappist-1f predicted atmosphere–interior time evolution from waterworld sensitivity test. In this calculation, only model runs that terminate with a water mass fraction >2.4% (and <14%) are plotted, representing a scenario where the planet's core mass fraction exceeds 25% (and is <50%), meaning a large surface volatile inventory is required to account for observed density. The lines are median values and shaded regions denote 95% confidence intervals. The time evolution spans from post-accretionary magma ocean to the present day ($8 \times 10^9$ yr). Subplot (a) shows the evolution of mantle potential temperature (magenta) and surface temperature (orange) alongside the silicate solidus (black dashed line). The magma ocean (b) persists for ∼100 Myr due to the large assumed initial water inventory. Water that degasses from the magma ocean is subsequently lost to space via hydrodynamic escape (c, j). The runaway greenhouse phase ends when Trappist-1 dims sufficiently for absorbed shortwave radiation to drop below the runaway greenhouse limit (e). When this occurs, all remaining atmospheric water vapor (c) condenses onto the surface producing an ocean (d). Subsequently, the overburden pressure from surface volatiles suppresses crustal oxygen sinks (f, i). Atmospheric oxygen produced via escape during the pre-main sequence (i, j) is more likely to persist over 8 Gyr (c).

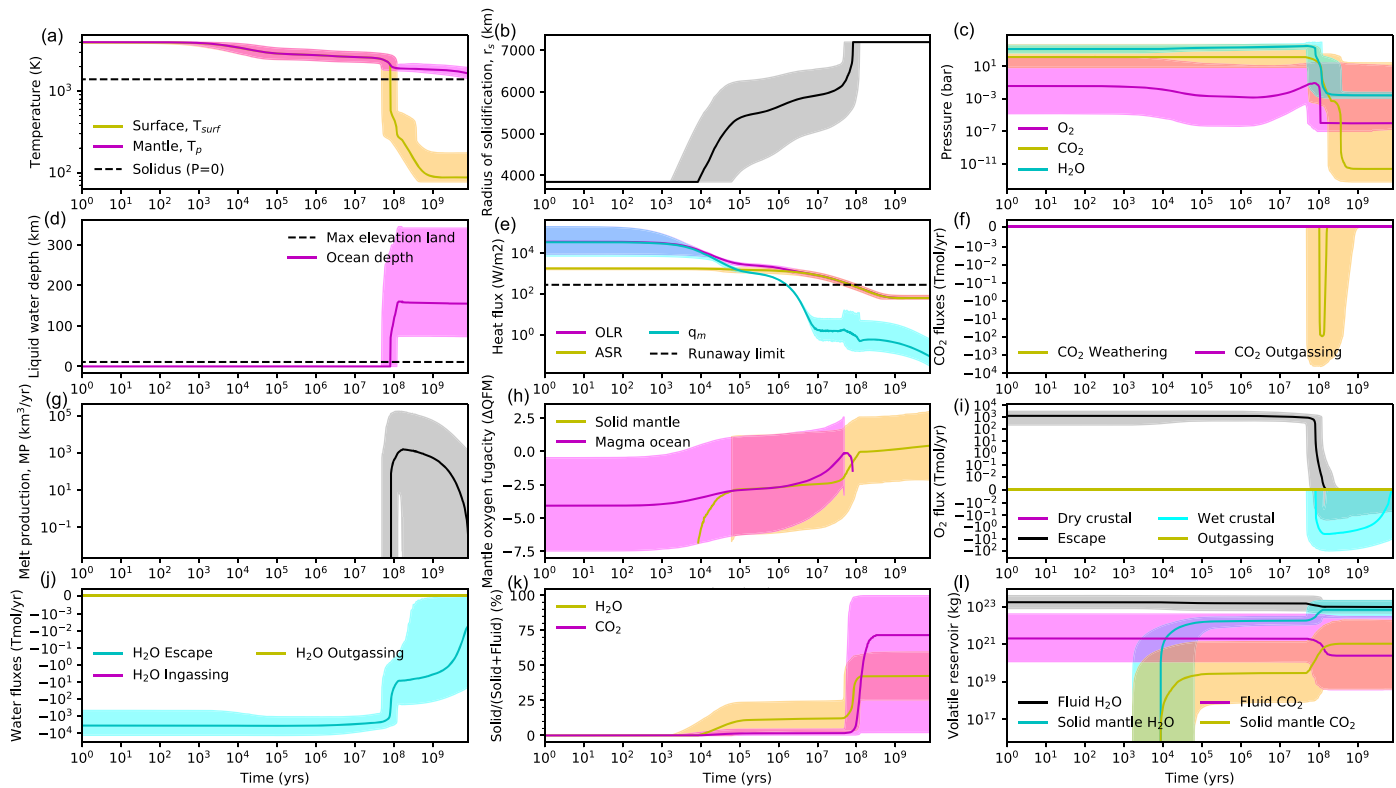

**Figure D6.** Trappist-1g predicted atmosphere–interior time evolution from waterworld sensitivity test. In this calculation, only model runs that terminate with a water mass fraction >0.57% (and <16%) are plotted, representing a scenario where the planet's core mass fraction exceeds 25% (and is <50%), meaning a large surface volatile inventory is required to account for observed density. The lines are median values and shaded regions denote 95% confidence intervals. The time evolution spans from post-accretionary magma ocean to the present day ($8 \times 10^9$ yr). Subplot (a) shows the evolution of mantle potential temperature (magenta) and surface temperature (orange) alongside the silicate solidus (black dashed line). The magma ocean (b) persists for ∼100 Myr due to the large assumed initial water inventory. Water that degasses from the magma ocean is subsequently lost to space via hydrodynamic escape (c, j). The runaway greenhouse phase ends when Trappist-1 dims sufficiently for absorbed shortwave radiation to drop below the runaway greenhouse limit (e). When this occurs, all remaining atmospheric water vapor (c) condenses onto the surface producing an ocean (d). Subsequently, the overburden pressure from surface volatiles suppresses crustal oxygen sinks (f, i), although the likelihood of oxygen retention after 8 Gyr is not dramatically different to the nominal model because initial oxygen accumulation is lower than 1e and 1f on account of orbital separation (shorter runaway greenhouse phase and less XUV flux to drive hydrogen escape).

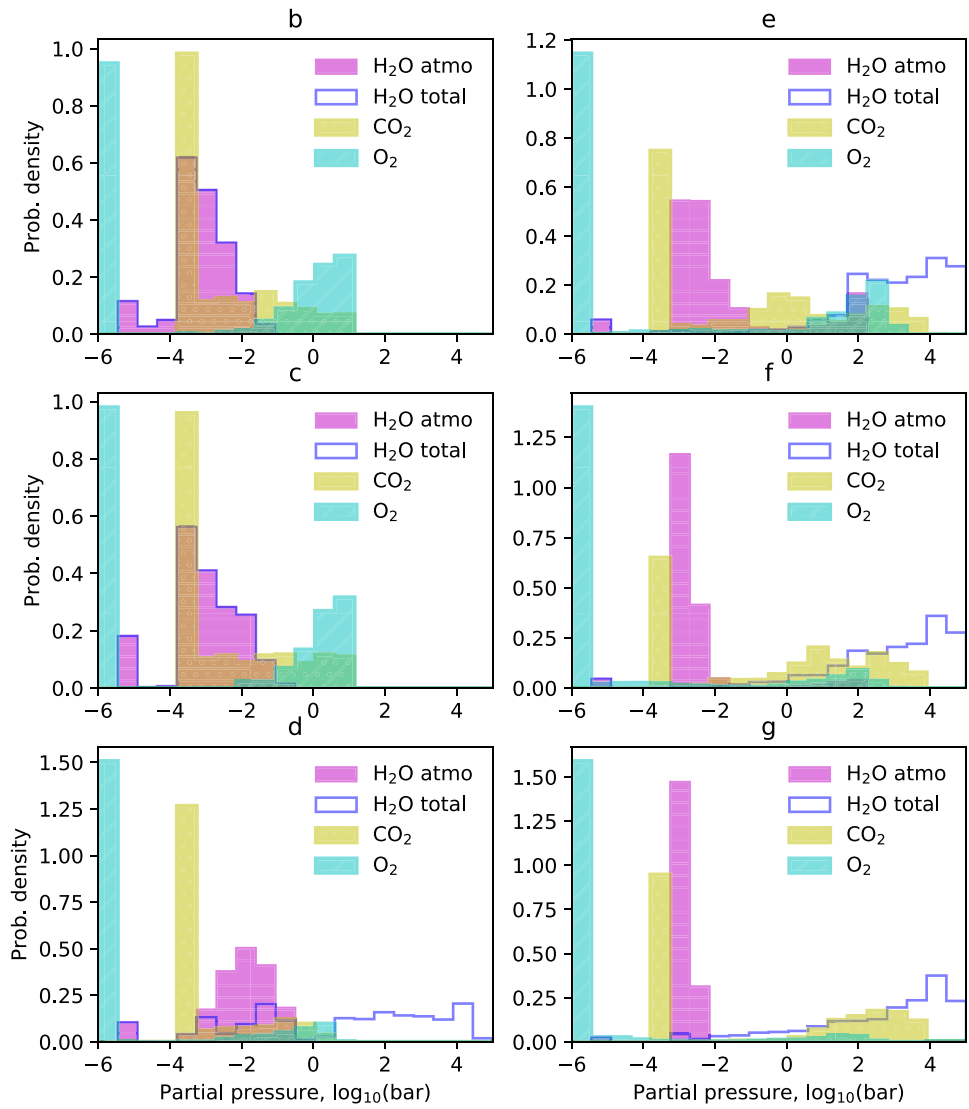

**Figure D7.** Probability distributions for modern atmospheric abundances of Trappist-1 planets for the Venus-like dry oxidation sink and smaller core mass fraction sensitivity test. Partial pressure distributions for carbon dioxide (yellow), oxygen (cyan), and water vapor (magenta) are shown for 1b, c, d, e, f, and g. The total surface water inventory is denoted by the unfilled, blue line distribution; this is identical to the water vapor distribution for the inner planets, but not for the outer planets because surface water can condense as a liquid or solid. Note that there are numerical cutoffs for carbon dioxide, oxygen, and water vapor at $10^{-4}$, $10^{-6}$, and $10^{-6}$, respectively.

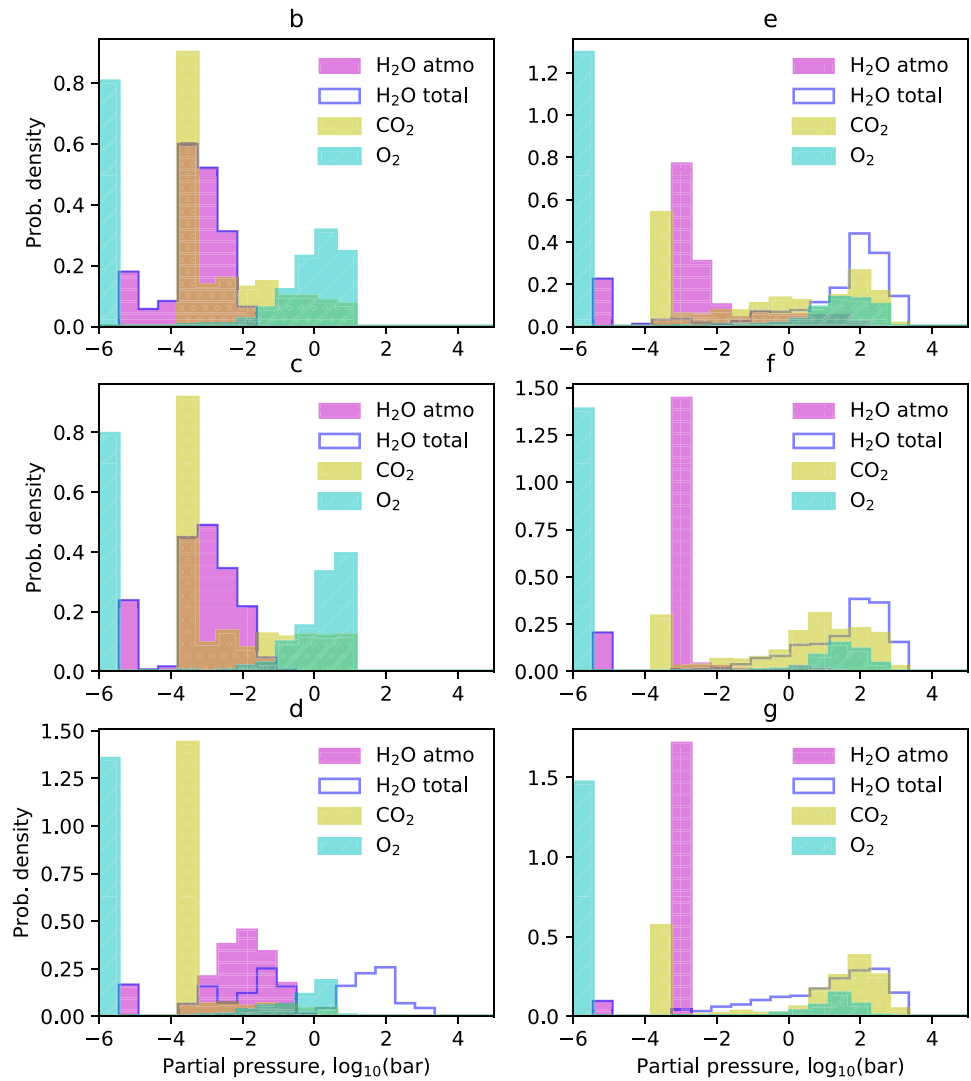

**Figure D8.** Probability distributions for modern atmospheric abundances of Trappist-1 planets for the Earth-like initial volatile inventory and smaller core mass fraction sensitivity test. Partial pressure distributions for carbon dioxide (yellow), oxygen (cyan), and water vapor (magenta) are shown for 1b, c, d, e, f, and g. The total surface water inventory is denoted by the unfilled, blue line distribution; this is identical to the water vapor distribution for the inner planets, but not for the outer planets because surface water can condense as a liquid or solid. Note that there are numerical cutoffs for carbon dioxide, oxygen, and water vapor at $10^{-4}$, $10^{-6}$, and $10^{-6}$, respectively.

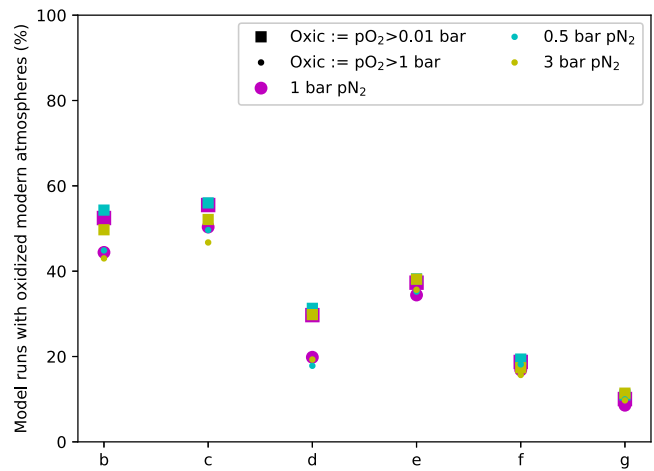

**Figure D9.** Percentage of model runs with oxygen-rich atmospheres after 8 Gyr of evolution for planets 1b–g, including sensitivity tests for different background $N_2$ partial pressures. Squares denote results when "oxygen-rich" is defined as $pO_2$ exceeding 0.01 bar, whereas circles denote a threshold of 1 bar. The latter is the approximate detectability threshold for JWST (Fauchez et al. 2020b). Different colors represent model runs from nominal 1 bar $N_2$ calculations (magenta), 3 bar background $N_2$ (yellow), 0.5 bar background $N_2$ (cyan).

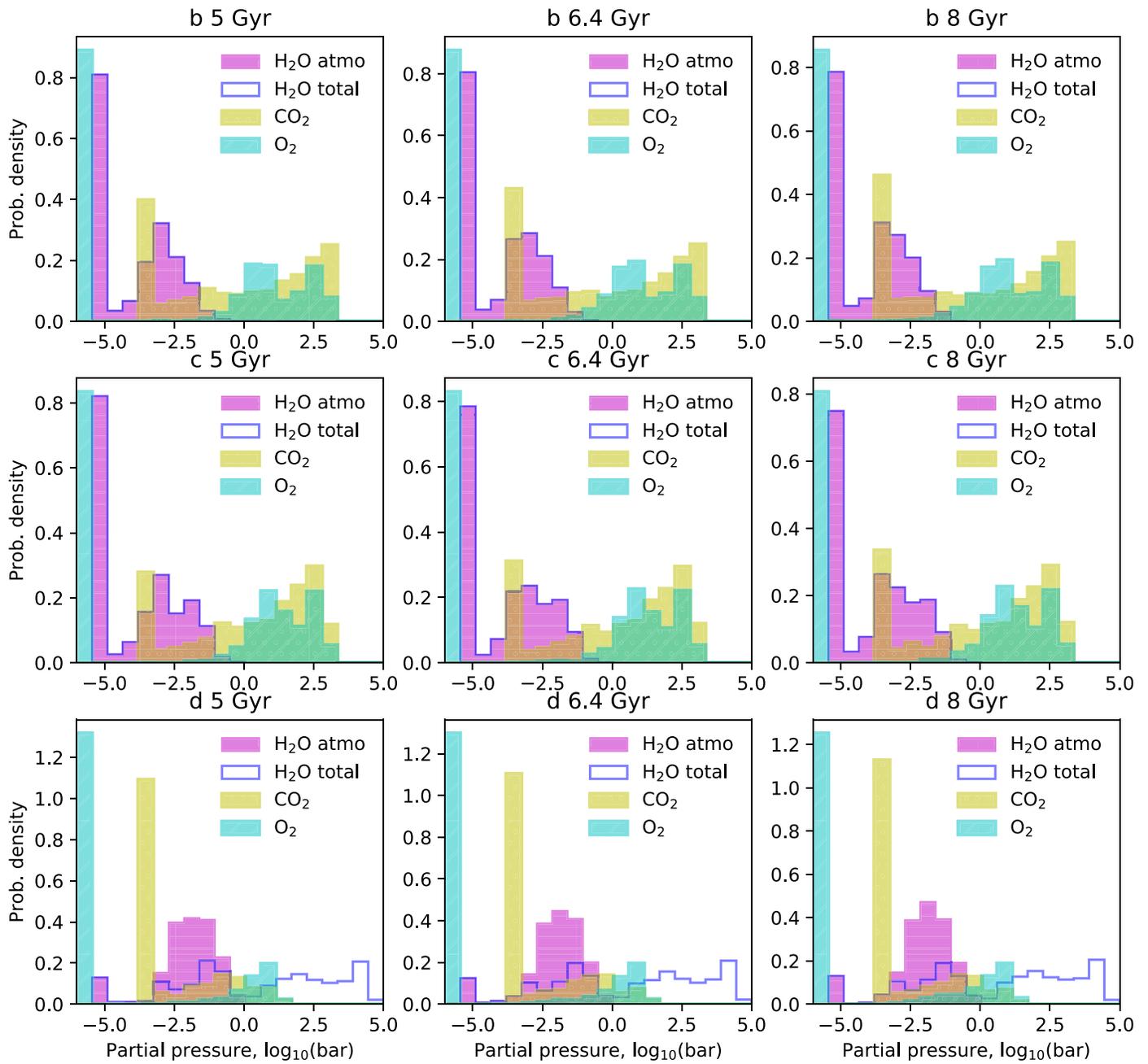

**Figure D10.** Current atmospheric composition of Trappist-1b (top), c (middle), and d (bottom) assuming a stellar age of 5 Gyr (left), 6.4 Gyr (middle), and 8 Gyr (right).

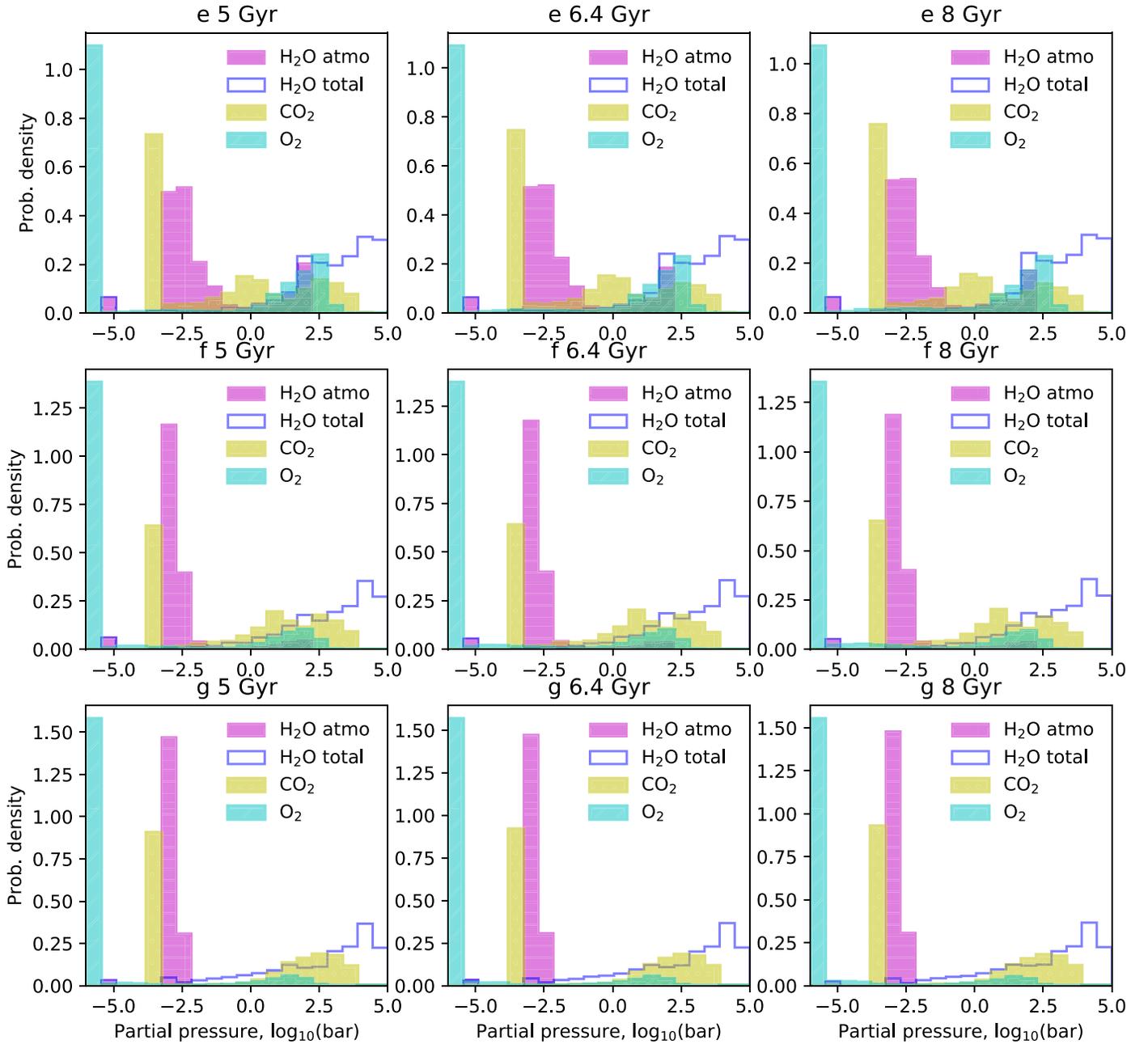

**Figure D11.** Current atmospheric composition of Trappist-1e (top), f (middle), and g (bottom) assuming a stellar age of 5 Gyr (left), 6.4 Gyr (middle), and 8 Gyr (right).

upper-atmosphere $H_2O$ mixing ratios, and therefore smaller diffusion-limited hydrogen escape rates as surface water inventories are diminished. For the other planets, there is little variation in oxygenation likelihood with changing $N_2$. Note that the effects of changing $N_2$ are not necessarily monotonic: more $N_2$ lowers the $H_2O$ mixing ratio for the same total water inventory, but more $N_2$ also warms the surface climate slightly, thereby resulting in more upper-atmosphere water vapor.

### D.6. Stellar Age

While stellar age does not correlate with any of the other stellar parameters in our Monte Carlo calculations (Section 2.2), it is conceivable that planetary evolution outcomes may depend on the current age of the star. For example, the decay of radiogenic elements that affect melt production could affect geochemical cycling, and therefore the observable atmosphere. Figures D10 and D11 show the present-day atmospheric composition of the Trappist-1 planets for three different stellar ages: 5 Gyr, 6.4 Gyr, and 8 Gyr. We find that the affect of stellar age on atmospheric composition is minimal.

## Appendix E
## Surface Temperature Distributions

Figure E1 shows the surface temperature probability distributions for the present-day Trappist-1 planets. These temperature distributions should be viewed as approximations since albedo is a randomly sampled parameter, not calculated self-consistently from atmospheric composition and surface coverage. Note that since approximately half of the model runs for the inner planets terminate with an airless rock, the surface temperature is often equal to the planetary equilibrium temperature (with a range determined by the sampled albedo range).

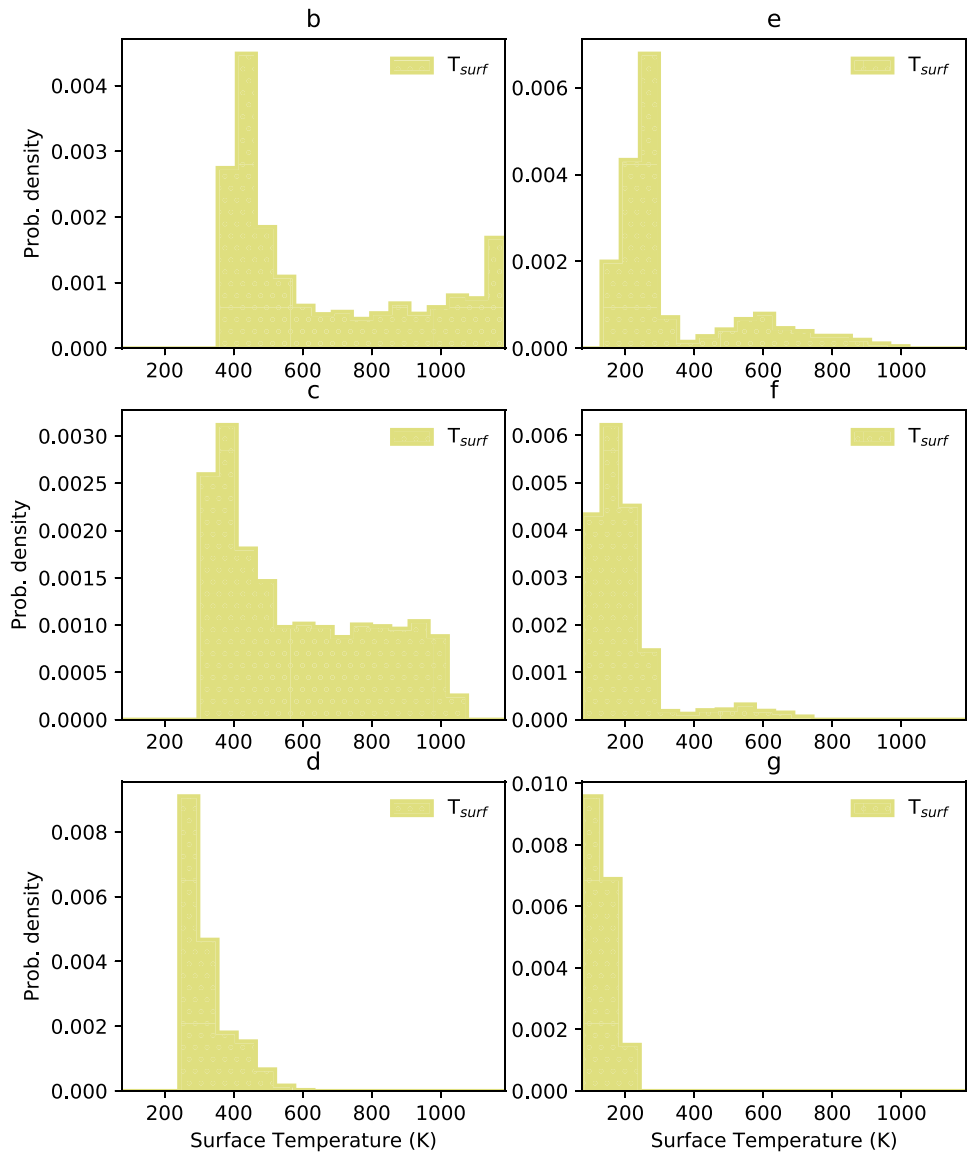

**Figure E1.** Present-day surface temperature distributions for Trappist-1b–g from nominal model.

## ORCID iDs

J. J. Fortney  https://orcid.org/0000-0002-9843-4354